\documentclass[11pt]{article}
\usepackage[a4paper,pdftex]{geometry}
\usepackage[T1]{fontenc}
\usepackage[utf8x]{inputenc}
\usepackage{amsmath,amsthm,amssymb,graphicx,enumerate,booktabs,bigstrut,rotating,multirow,float,caption,etoolbox,hyperref,lmodern,comment,listings,qtree,a4wide,titlesec,moresize,bbm,subcaption,tikz,pdfescape,mathtools,flafter,enumitem,setspace,ragged2e,colortbl,lineno,lscape}
\hypersetup{colorlinks,linkcolor={red},citecolor={blue},urlcolor={blue}}  

\usepackage[justification=centering,font=small]{caption}

\usepackage[authoryear]{natbib}
\usepackage[english]{babel}

\makeatletter
\renewcommand\subsubsection{\@startsection{subsubsection}{3}{\z@}%
	{-3.25ex\@plus -1ex \@minus -.2ex}%
	{-1.5ex \@plus -.2ex}
	{\normalfont\normalsize\bfseries}}
\def\@biblabel#1{\hspace*{-\labelsep}}

\makeatother

\def\sym#1{\ifmmode^{#1}\else\(^{#1}\)\fi}
\pdfpageheight\paperheight
\pdfpagewidth\paperwidth

\newcolumntype{L}[1]{>{\raggedright\let\newline\\\arraybackslash\hspace{0pt}}m{#1}}
\newcolumntype{C}[1]{>{\centering\let\newline\\\arraybackslash\hspace{0pt}}m{#1}}
\newcolumntype{R}[1]{>{\raggedleft\let\newline\\\arraybackslash\hspace{0pt}}m{#1}}

\begin{document}
	
	\title{Are Fairness Perceptions Shaped by Income Inequality? Evidence from Latin America\thanks{\noindent Gasparini: Centro de Estudios Distributivos, Laborales y Sociales (CEDLAS) - IIE-FCE, Universidad Nacional de La Plata and CONICET (e-mail: \text{lgasparini@cedlas.org}). Reyes: Cornell University (e-mail \text{gjr66@cornell.edu}). For helpful comments, we thank Carolina Garc\'ia Domench, Giselle Del Carmen, Rebecca Deranian, Luis Laguinge, participants of the 2021 Annual Bank Conference on Development Economics, and two anonymous referees. A previous version of this article circulated as ``Perceptions of Distributive Justice in Latin America during a Period of Falling Inequality.'' Errors and omissions are our own.}} 
	
	\author{Leonardo Gasparini \and Germ\'an Reyes}
	
	\date{\vspace{15pt} February 2022}
	
	\maketitle
	
	\begin{abstract}
		
		\noindent A common assumption in the literature is that the level of income inequality shapes individuals' beliefs about whether the income distribution is fair (``fairness views,'' for short). However, individuals do not directly observe income inequality (which often leads to large misperceptions), nor do they consider all inequities to be unfair. In this paper, we empirically assess the link between objective measures of income inequality and fairness views in a context of high but decreasing income inequality. We combine opinion poll data with harmonized data from household surveys of 18 Latin American countries from 1997--2015. We report three main findings. First, we find a strong and statistically significant relationship between income inequality and unfairness views across countries and over time. Unfairness views evolved in the same direction as income inequality for 17 out of the 18 countries in our sample. Second, individuals who are older, unemployed, and left-wing are, on average, more likely to perceive the income distribution as very unfair. Third, fairness views and income inequality have predictive power for individuals' self-reported propensity to mobilize and protest independent of each other, suggesting that these two variables capture different channels through which changes in the income distribution can affect social unrest. \\
		
		
		
	\end{abstract}
	
	\clearpage

		\section{Introduction}

Several theoretical and empirical papers study how objective measures of income inequality affect individual-level behavior.\footnote{Researchers have studied how income inequality affects cooperation \citep{cozzolino2011trust}, demand for redistribution \citep{meltzer1981rational, finseraas2009income}, dishonesty \citep{neville2012economic}, social cohesion \citep{alesina1996income}, subjective wellbeing \citep{oishi2011income}, and trust \citep{gustavsson2008inequality}.} Implicit in much of this literature is the assumption that inequality shapes individuals' beliefs about whether the income distribution is fair (``fairness views,'' for short).\footnote{For example, one important reason why social cohesion might be related to income inequality is that, as inequality increases, more individuals perceive the income distribution as unfair, making them more prone to mobilize. Similarly, we would not expect inequality to be negatively linked to subjective wellbeing if increases in income disparities were perceived as fair.} However, two pieces of evidence suggest that the link between inequality and fairness views is not straightforward. First, fairness views are not informed by objective measures of inequality---since these are not directly observable by individuals---but instead by perceived inequality. Research about how accurately people perceive income inequality shows large gaps between individuals' perceptions and the actual levels of inequality \citep{norton2011building,kuziemko2015elastic, gimpelson2018misperceiving, choi2019revisiting}. Second, even absent any misperceptions, individuals do not consider all inequities to be unfair. Specifically, individuals largely accept income disparities derived from personal choices and effort, but deem inequities driven by luck or chance as unfair \citep{cappelen2007pluralism, alesina2011preferences, cappelen2013just, almaas2020cutthroat}. Thus, the extent to which fairness perceptions are shaped by income inequality remains an important empirical question.

In this paper, we empirically study the link between fairness views and income inequality in a particular scenario: a region of highly unequal countries---Latin America---but during a period in which inequality pronouncedly declined. First, we assess the extent to which fairness views are linked to income inequality, both across countries and over time. Then, we analyze how individual-level factors such as education, political ideology, and religious views relate to fairness views. Finally, we investigate the predictive power of fairness views for individuals' propensity to protest and mobilize.

In Section \ref{sec_context_data}, we describe the institutional context and our data. Our setting is Latin America, one of the most unequal regions in the world \citep{alvaredo2015recent}. We focus on the 2000s, an unusual period in that income inequality saw a widespread decrease across countries in the region \citep{gasparini2011recent}. To relate income inequality to fairness views, we combine data from two harmonization projects. Our source for income inequality data is SEDLAC, a project that increases cross-country comparability of household surveys. These data enable us to compare the evolution of income inequality across countries and over time. The data on fairness perceptions come from public opinion polls conducted by Latinobar\'ometro in 18 Latin American countries since the 1990s. 

In Section \ref{sec_ineq_fairness}, we document a series of stylized facts about fairness views in the region. A strikingly high, albeit decreasing, share of the population believes that the income distribution in their country is unfair. In 2002, almost nine out of ten individuals perceived the income distribution as unfair (86.6\%). By 2015, this figure declined to 75.1\%. This decline is particularly striking considering that previous research highlights people's tendency to report perceptions of stable or increasing inequality, regardless of its actual evolution \citep{gimpelson2018misperceiving}. 

Next, we link fairness views to income inequality. We study how the Gini coefficient---our main measure of inequality---correlates with fairness views across countries and over time. We find a strong linear correlation between the Gini coefficient and the percent of the population that perceives inequality as unfair across country-years. Fairness views evolved in the same direction as the Gini in 17 out of the 18 countries in our sample during the 2000s. 

The decline in income inequality during the 2000s---although remarkable by historical standards---was not enough to substantially modify the view of Latin Americans with respect to distributive fairness in their societies. Three out of four citizens of the region still believe that the income distribution is unfair. A one percentage point decrease in the Gini is associated with a 1.4 percentage point decrease in the share of the population perceiving the income distribution as unfair. Holding constant this elasticity and the pace of inequality reduction of the 2000s, reducing the population that perceives inequality as unfair to 50\% would take roughly more than a decade.

Next, we investigate whether inequality measures other than the Gini coefficient are stronger predictors of unfairness beliefs. This question is of interest in its own right, given the ongoing debate on whether income inequality should be measured with relative or absolute indicators \citep{ravallion2003debate, atkinson2010analyzing}. We take an agnostic approach and correlate a large number of relative and absolute measures of inequality with unfairness views. We find that relative indicators are strongly positively correlated with people's perceptions of unfairness. In contrast, absolute indicators tend to be \textit{negatively} correlated with unfairness views. This is because absolute income gaps became wider in Latin America in the booming years under analysis, yet perceptions about unfairness went down.

In Section \ref{sec_corr_fairness}, we assess whether the correlation between income inequality and fairness views is robust to controlling for observable variables and investigate which individual-level characteristics are predictive of fairness views. The relationship between unfairness views and the Gini coefficient is positive and statistically significant even after controlling for country fixed effects, year fixed effects, and a large set of individual-level characteristics. We find that older, unemployed, and left-wing individuals are more likely to perceive income distribution as very unfair. A decomposition exercise shows that the decline in unfairness perceptions during the 2000s is better accounted for by income inequality trends rather than changes in the composition of the population.

In Section \ref{sec_unrest}, we analyze the link between fairness perceptions and individuals' self-reported propensity to protest. A vast literature relates income inequality to social cohesion, conflict, and activism. One might expect this link to be partly mediated by fairness views. Hence, we study whether fairness views have predictive power for social unrest, conditional on income inequality. To do this, we measure individuals' likelihood of participating in different political activities, such as mobilizing in a demonstration, signing a petition, refusing to pay taxes, or complaining on social media. Participation in these activities is self-reported, so the results should be interpreted with caution. With this caveat in mind, we find that both fairness views and inequality have predictive power independent of each other for some political activities, such as complaining on social media. Other behaviors are exclusively predicted by fairness views (e.g., signing a petition) or by income inequality (e.g., refusing to pay taxes). This suggests that both fairness views and income inequality are important determinants of the propensity to engage in political activism.

This paper contributes to the literature that links objective measures of the income distribution to individuals' perceptions of such measures. Previous papers have shown that individuals tend to misperceive their relative incomes \citep{cruces2013biased, karadja2017richer, hvidberg2020social, fehr2021your} and other relevant features of the income distribution, such as the level of inequality, poverty, and mobility \citep{kuziemko2015elastic, page2016subjective, alesina2018intergenerational, preuss2020}.\footnote{Mismatches between beliefs and reality are important because there is mounting evidence that perceptions of facts, more than facts themselves, affect individual behavior. For example, demand for redistribution is affected more by perceptions of the income distribution than by the actual distribution \citep{gimpelson2018misperceiving, choi2019revisiting}. Thus, understanding what people believe about the income distribution is important from a policy perspective. If there are mismatches between perceptions and reality, interventions that make information less costly or more salient might be desirable.} Evidence on the relationship between fairness perceptions and income inequality, particularly in Latin America, is rather scarce. To our knowledge, the only other paper that studies the link between inequality and fairness views in the region is \cite{zmerli2015income}, although the focus of this paper is on political trust. Using the same data that we use, the authors find a positive association between unfairness views and the Gini. However, the authors only use data from one year. Hence, they cannot study the joint evolution of both variables over time or control for unobserved heterogeneity at the country or year level, which, as we argue below, could generate a spurious correlation between income inequality and fairness views. We contribute to this literature by providing novel empirical evidence linking fairness views to income inequality in a highly unequal region, but during a period of falling inequality.\footnote{A related literature exploits opinion surveys to study distributive issues in Latin America. \cite{cepal2010america} documents patterns of perceptions of distributive inequity during 1997--2007. Using data from Argentina, \cite{rodriguez2014percepciones} shows that people who consider their income to be fair tend to perceive lower levels of inequality. \cite{martinez2020latin} explore the effect of immigration on preferences for redistribution in Latin America.}

This paper also contributes to the literature on inequality measurement. This literature makes a crucial distinction between two types of inequality indicators: the relative ones (such as the Gini coefficient) and absolute ones (such as the variance). Relative and absolute indicators often provide different answers to important issues such as the distributive effects of globalization or trade openness \citep{ravallion2003debate, atkinson2010analyzing}. Hence, it is important to understand whether people think about distributive fairness through the lens of relative or absolute indicators. We show that relative indicators have a much stronger correlation with fairness views than absolute indicators.

Finally, we make a small contribution to the growing literature that relates income inequality---and more recently, measures of polarization---to conflict and political activism \citep{esteban2011linking, esteban2012ethnicity}. Previous papers have shown that income inequality is predictive of conflict and social unrest. We contribute to this literature by showing that fairness views have predictive power for social unrest above and beyond income inequality (and vice-versa).


	\section{Institutional Context, Data, and Descriptive Statistics} \label{sec_context_data}

This section provides institutional context on Latin America, describes our data, and provides summary statistics on our sample.

\subsection{Context} \label{subsec_context}

Latin America has long been characterized as a region with high levels of income inequality. Together with South-Saharan Africa, Latin America is one of the two most unequal regions in the world \citep{alvaredo2015recent,world2016poverty}. After a period of increasing inequality during the 1980s and 1990s, the region experienced a ``turning point'' in the 2000s, when income inequality saw a widespread decrease \citep{gasparini2011recent, gasparini2011rise, lustig2013declining}.\footnote{The widespread decline in inequality contrasts to what happened in other developing regions in the world, where inequality modestly decreased (e.g., such as in the Middle East and North Africa), or even increased (such as in East Asia and Pacific, \citealp[c.f.][]{alvaredo2015recent}). In developed countries, inequality tended to increase \citep{atkinson2011top}.}

\subsection{Data} \label{subsec_data}

We use data on fairness views and income inequality from 18 Latin American countries from 1997--2015. The data comes from two harmonization projects, known as Latinobar\'ometro and SEDLAC (Socio-Economic Database for Latin America and the Caribbean).

We use public opinion polls conducted by Latinobar\'ometro to measure fairness perceptions. Latinobar\'ometro conducts opinion surveys in Latin American countries, interviewing about 1,200 individuals per country. The survey is designed to be representative of the voting-age population at the national level (in most Latin American countries, individuals aged over 18).\footnote{In Appendix Table \ref{tab-coverage} we show the percentage of the voting-age population represented by the opinion polls in our sample for the years in which fairness data is available.} The main variable for our empirical analysis is individuals' fairness views, which we measure using the following question: \textit{``How fair do you think the income distribution is in [country]? Very fair, fair, unfair or very unfair?''} We construct binary variables indicating whether individuals believe that the income distribution is unfair or very unfair.\footnote{Latinobar\'ometro does not ask respondent \textit{why} they believe that the income distribution is unfair. It is possible that some people view the distribution as unfair because  existing disparities are not sufficiently large. We think that this is unlikely and interpret unfairness views as reflective of too much inequality.}

Income inequality data comes from SEDLAC, a joint project between CEDLAS-UNLP and The World Bank, which increases cross-country comparability from official household surveys. We measure inequality in household income per capita (measured in 2005 USD at purchasing power parity). Whenever possible, we use comparable annual household surveys to calculate inequality indicators. However, some countries do not conduct surveys every year, and some of the household surveys available in a given country are not comparable over time (usually, due to important methodological changes). In Appendix \ref{sec_data} we describe the partial fixes we implement to maximize the sample size. In two countries of our sample (Argentina and Uruguay), the household survey is representative of urban areas only.

\subsection{Sample and Summary Statistics} \label{subsec_sample}

For our regression analysis, we use individual-level data. Our sample includes all individuals in the 11 different waves of Latinobar\'ometro surveys over 1997--2015.

Appendix Tables \ref{tab-sum-stats} and \ref{tab-fairness-grp} show descriptive statistics on our sample. The average respondent is 39.7 years old. Roughly half of the respondents are men (49\%), over half (56.3\%) are married or in a civil union, and 68\% are Catholics. The majority of respondents (76\%) completed at least elementary school, while a third of them (33.6\%) had completed high school or more. Almost two-thirds of the sample (64\%) were part of the labor force, and 9.9\% of the economically active individuals were unemployed.\footnote{In Appendix \ref{app_lb_sedlac}, we show that Latinobar\'ometro's sample is similar to the sample in SEDLAC.} 

	\section{The Relationship between Fairness Views and Inequality} \label{sec_ineq_fairness}

This section first provides descriptive evidence on the evolution of fairness views in Latin America from 1997 to 2015. Then, we link fairness perceptions to income inequality both across countries and over time. Finally, we investigate whether absolute or relative inequality indicators have a stronger predictive power for fairness views.

\subsection{The Evolution of Fairness Views in Latin America during the 2000s}

\begin{figure}[htp]{\scriptsize
		\begin{centering}	
			\protect\caption{Fairness views in Latin America over time and across countries} \label{fig-fairness-views}
			\begin{minipage}{.48\textwidth}
				\caption*{Panel A. Fairness views over time \\ (1997--2015)}\label{fig-fairness-views-time}
				\includegraphics[width=\linewidth]{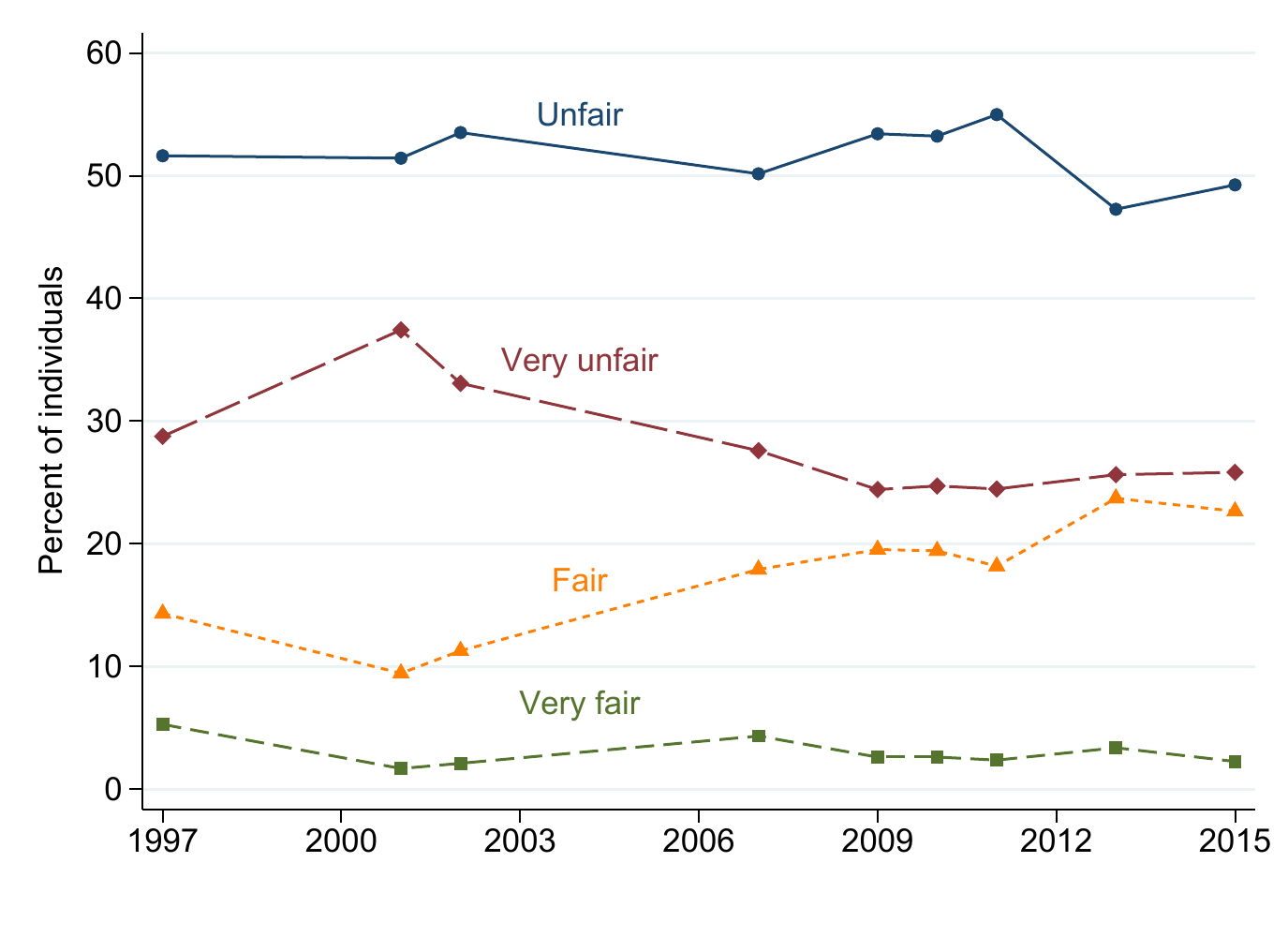}
			\end{minipage}\hspace{1em}
			\begin{minipage}{.48\textwidth}
				\caption*{Panel B. Fairness views across countries \\ (2003 vs. 2013)} \label{fig-fairness-views-countries}				
				\includegraphics[width=\linewidth]{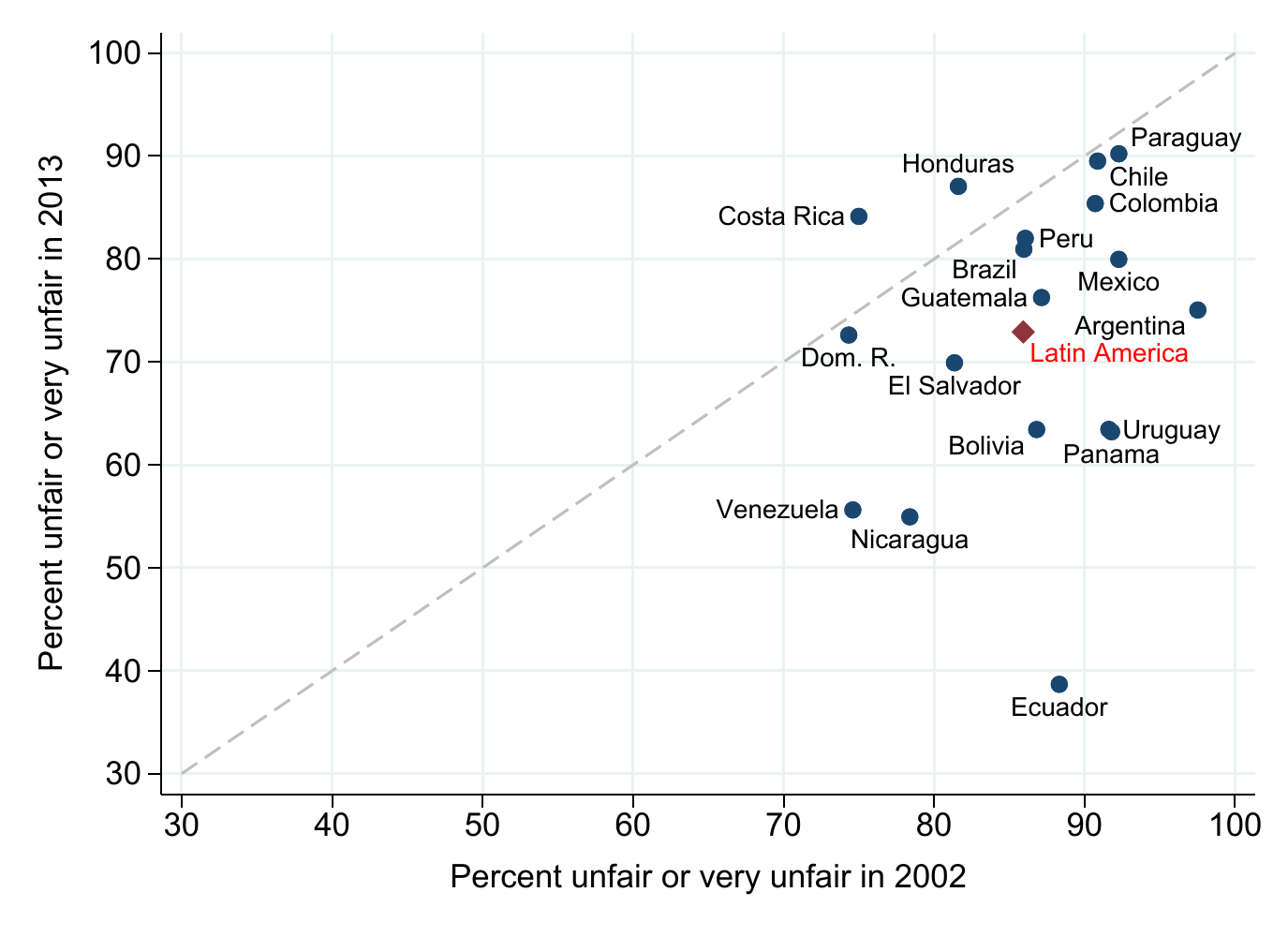}
			\end{minipage}
			\par\end{centering}
		
		\singlespacing\justify
		\textbf{Notes}: \footnotesize Panel A shows the percentage of individuals who perceive the income distribution as very unfair, unfair, fair, and very fair in our sample. We calculate the shares as the unweighted average of fairness views across the 18 countries in our sample. Panel B presents the percentage of the population who perceives the income distribution as either unfair or very unfair in 2002 and 2013 in each country of our sample.
		
	}
\end{figure}

Figure \ref{fig-fairness-views} shows how fairness views evolved over time (Panel A) and across countries (Panel B). Panel A plots the fraction of individuals who believe that the income distribution of their country is very unfair, unfair, fair, or very fair over 1997--2015, pooling across all countries in our sample. Panel B shows the fraction of individuals who believe that the income distribution is either unfair or very unfair in each country of our sample during 2002 and 2015.

Figure \ref{fig-fairness-views}, Panel A shows that a strikingly high, albeit decreasing, share of the population believes that the distribution of income is unfair. In 2002, almost nine out of ten individuals perceived the income distribution as unfair (86.6\%). By 2015, this figure declined to 75.1\%. The decrease in unfairness perceptions was driven mainly by individuals with strong beliefs about unfairness (i.e., individuals who believe that the income distribution was very unfair). While in 2002, 33.1\% of the population perceived the income distribution as very unfair, this figure declined to 25.8\% by 2015. In contrast, weak beliefs about unfairness (i.e., individuals who believe that the income distribution was merely unfair) behaved more erratically, increasing at the beginning of our sample and decreasing by the end of the period. Overall, the share of individuals with weak beliefs about unfairness slightly declined during the 2000s, from 53.5\% in 2002 to 49.2\% in 2015. On the other hand, the share of the population that believe that the income distribution is fair doubled from 11.3\% in 2001 to 22.6\% in 2015.

Figure \ref{fig-fairness-views}, Panel B shows that, while most individuals perceive the income distribution as unfair, fairness perceptions improved in most countries during 2002--2013. A substantial share of the population in all countries perceived the income distribution as unfair in both 2002 and 2013. For example, in 2002, the share of the population that perceived the distribution as unfair ranged from 74.5\% in Venezuela to 97.7\% in Argentina (which, at the time, was in the midst of a severe economic crisis). Throughout the following decade, there was a widespread decrease in the share of the population that perceived income inequality as unfair or very unfair. Compared to 2002, in 2013, a lower fraction of the population perceived the income distribution as unfair in 16 out of the 18 countries in our sample. The change in fairness perceptions ranged from modest decreases, like in Chile, where the decline was of less than one percentage point, to remarkable reductions, like in Ecuador, where perceptions about unfairness declined from 87.5\% to 38.6\%.

Appendix Figure \ref{fig-fairness-grps} shows that the decline in unfairness perceptions was widespread across heterogeneous groups of the population. To show this, we study how fairness views evolved for different subgroups of the population, according to individuals’ age, gender, education, and employment. This analysis reveals that young individuals are less likely to perceive the income distribution as unfair (Panel A), while females are more likely to do so, although with some heterogeneity across time (Panel B). Similarly, individuals with a higher educational achievement (Panel C) and the unemployed (Panel D) are more likely to believe that the income distribution is unfair. Importantly, the perception of unfairness consistently fell across all these subpopulations during the 2000s.

Next, we explore the extent to which these changes in fairness views were accompanied by changes in the actual distribution of income.

\subsection{Fairness Perceptions and Income Inequality: Some Stylized Facts}

Figure \ref{fig-fairness-gini} shows how fairness views evolved vis-\`a-vis changes in income inequality. Panel A shows a binned scatterplot of the Gini coefficients and unfairness views for all country-years in our sample. Panel B plots the percentage point change in unfairness perceptions between 2002 and 2013 on the $y$-axis against the change in the Gini over the same period on the $x$-axis. 

Panel A shows that unfairness perceptions and income inequality are strongly correlated across country-years. The linear correlation between the Gini and unfairness perceptions across country-years is $0.93$ ($p < 0.01$). The share of the population who perceive income as unfair or very unfair ranges from 63\% in country-years with a Gini in the 0.40 bin (roughly, the average level of inequality in Venezuela during the 2000s), to 88\% in country-years with a Gini in the 0.60 bin (roughly, the level of inequality in Honduras during the early 2000s).\footnote{An OLS regression of unfairness views on the Gini estimated on the plotted points yields an intercept of 28.2. This implies that, even in a society where all incomes are equalized, about 28\% of the population would still perceive the income distribution as unfair. This exercise relies on the strong assumption that the relationship between fairness views and income inequality is linear. While such a relationship indeed appears to be linear in Panel A, our data only covers a very narrow range of Gini coefficients (between 0.40 and 0.60). It is likely that the relationship is non-linear for Gini coefficients close to zero or one.} The correlation between unfairness views and the Gini is driven by individuals who perceive inequality as very unfair. The correlation between perceptions of a very unfair distribution and the Gini is sizable and statistically significant. In contrast, the correlation between perceptions of a merely unfair distribution and the Gini is small and indistinguishable from zero.

Panel B shows that the evolution of fairness views tends to mirror the evolution of income inequality at country level. Fairness views moved in the same direction as the Gini in 17 out of the 18 countries in our sample. The one exception is Honduras, where, despite falling inequality, the population perceived the distribution as more unjust. Most countries lie in the third quadrant, where both the Gini and unfairness perceptions decreased. The only country where inequality increased (Costa Rica), also saw an increase in unfairness beliefs. In Appendix Figure \ref{fig-timeseries-gini-unfair} we show that the correlation between the Gini and unfairness views over time is also strong when pooling across countries. In this case, the linear correlation is equal to 0.80 ($p < 0.01$). During 2002--2013, a one percentage point decrease in the Gini was associated with a 1.4 percentage point decrease in the share of the population perceiving the distribution as unfair. To put this figure in context, this means that, at the pace of inequality reduction of the 2000s, it would roughly take Latin America more than a decade to reduce the population that perceives income inequality as unfair to 50\%.

\subsection{Is Fairness Absolute or Relative?} \label{sec_abs_fairness}


The literature on inequality measurement makes a crucial distinction between two types of indicators: relative ones (such as the Gini) and absolute ones (such as the variance). Relative indicators fulfill the scale-invariant axiom, while the absolute indicators meet the translation-invariant axiom.\footnote{These two axioms yield different implications for how inequality responds to a proportional change in the income of the entire population. A proportional income increase does not generate changes in income inequality as measured by relative indicators, but can provoke a large increase in inequality as measured by absolute indicators.} The question of which indicator should be used in practice has led to a heated debate in the literature \citep{milanovic2016global}. This is because relative and absolute indicators often provide different answers to important issues such as the distributive effects of globalization or trade openness.\footnote{As measured by absolute indicators, globalization has deteriorated the income distribution since the absolute income difference between the rich and the poor has increased. However, under the lens of relative measurement, globalization reduced income inequality since the poor's income has grown proportionally more than the income of the rich.} 

To shed some light on this debate, we assess whether people think about distributive fairness through the lens of relative or absolute indicators. To do this, we take a data-driven approach. We calculate 13 different measures of income inequality for all the countries in our sample and correlate each inequality indicator with the share of the population that believes income distribution is unfair over time.\footnote{The indicators are the Gini coefficient, the ratio between the 75th percentile and the 25th percentile of the income distribution, the ratio between the 90th and 10th percentile, the Atkinson index with an inequality aversion parameter equal to 0.5 and 1, the mean log deviation, the Theil index, the Generalized entropy index, the coefficient of variation, the absolute Gini, the Kolm index with an inequity aversion parameter equal to one, and the variance of the per capita household income (in 2005 PPP). These last three indices correspond to the absolute inequality measures, while the other ten indicators are relative inequality measures.} 

We calculate the correlation between unfairness perceptions and each inequality indicator at the regional level using three alternative aggregation methods. First, we calculate each correlation using the individual-level data and pooling all countries and years in our sample (columns 1-3). Second, we calculate the average unfairness views in each country-year and then calculate the correlation between each inequality indicator and the average fairness views in the corresponding country-year (columns 4-6). Third, we calculate the correlation between each inequality indicator and fairness views over time for each country separately (using the individual-level data) and then average the correlations across countries (columns 7-9). Table \ref{tab-rel-fairness} shows the results.

Fairness views tend to be positively correlated with relative indicators and negatively correlated with absolute ones. In Table \ref{tab-rel-fairness}, column 1 shows that the Gini is the indicator with the highest linear correlation. On the other hand, the absolute indicators of inequality tend to be \textit{negatively} correlated with unfairness perceptions, and the magnitude of such correlations tends to be small. The high correlation between unfairness perceptions and income inequality seems to be driven by the population that perceives inequality as very unfair (columns 2, 5, and 8), rather than just unfair (columns 3, 6, and 9).\footnote{It is interesting to note that indicators sometimes mentioned in the mass media, such as the ratio between the richest 90\% and the poorest 10\%, exhibit low explanatory power. This may be due to mismeasurement of top incomes in household surveys.}

These results are consistent with experimental evidence from \cite{amiel1992measurement,amiel1999thinking}, who show that support for the scale-invariance axiom is greater than for translation invariance, reflecting greater support for relative inequality indicators. 

	\section{Individual-level Fairness Determinants} \label{sec_corr_fairness}

In this section, we study the individual-level correlates of fairness views. The purpose of this section is twofold. First, to assess whether the association between fairness views and income inequality is robust to including controls. Second, to investigate which individual-level characteristics are systematically related to fairness views. 

\subsection{Empirical Design}

To assess the relationship between individuals' characteristics and fairness perceptions, we estimate two-way fixed effects Logit models. This design controls for two important sources of heterogeneity that could drive the positive association between inequality and fairness perceptions documented in the previous section. First, it controls for country-level heterogeneity. This could matter if, for example, countries with historically extractive institutions have both higher levels of income inequality and more negative fairness views as a legacy of such institutions. Insofar as institutions are stable over time, the country fixed effects deal with this potential bias. Second, the design controls for year-level heterogeneity. This is important if, for example, in some particular years, macroeconomic events such as a financial crisis or a corruption scandal increase income disparities and worsen fairness views, again generating a spurious correlation between inequality and fairness perceptions. Including year fixed effects helps to alleviate such concerns.

Given that changes in fairness views over the last decade were driven by the share of the population that perceived the income distribution as very unfair (Figure \ref{fig-fairness-views}), in our baseline specification, we focus on explaining the determinants of this variable, although we also show the results for a broader definition of unfairness. We assume that unfairness perceptions can be characterized according to the following equation:
\begin{align} \label{eq_fairness}
	\text{VeryUnfair}_{ict} = F(\lambda_c + \lambda_t + \gamma \text{Gini}_{ct} + \beta x_{ict}),
\end{align}
where the dependent variable, $\text{VeryUnfair}_{ict}$ is equal to one if individual $i$ believes that the income distribution of country $c$ during year $t$ is very unfair and zero otherwise. Equation \eqref{eq_fairness} includes country fixed effects, $\lambda_{c}$; year fixed effects, $\lambda_t$; the country's Gini coefficient in year $t$, $\text{Gini}_{ct}$; and a vector of individual characteristics, $x_{ict}$, that contains age, age squared, sex, marital status, education, employment status, an assets index, political ideology, and religious views.\footnote{The assets index takes the value one if individual $i$ has access to running water and sewerage, owns a computer, a washing machine, a telephone, and a car. In household surveys, these variables tend to be correlated with household income, although the correlation is usually small. Unfortunately, we do not observe household income in the Latinobar\'ometro data. To measure political views, we rely on the question \textit{``In politics, people normally speak of ``left'' and ``right.'' On a scale where 0 is left and 10 is right, where would you place yourself?''} We interpret values closer to zero (ten) as closer to a liberal (conservative) worldview. We measure religious views using a dummy that is equal to one if individual $i$ is a Catholic ---the predominant religion in the region---coding other religions (and lack of religion) as zero.} In our baseline specification, $F(\cdot)$ is the logistic function. We cluster the standard errors at the country-by-year level.

We are interested in $\frac{\partial \text{VeryUnfair}_{ict}}{\partial x_{ict}} = \beta f(\cdot)$ and $\frac{\partial \text{VeryUnfair}_{ict}}{\partial \text{Gini}_{ict}} = \gamma f(\cdot)$. The first of these partial derivatives captures the relationship between an individual characteristic and unfairness perceptions, controlling for the rest of the characteristics, the Gini, and the fixed effects. Similarly, $\gamma f(\cdot)$ measures the relationship between the Gini and perceived fairness after controlling for individual-level traits and fixed effects. The magnitude of the partial derivatives depends on the value at which covariates are evaluated. We compute the marginal effects by evaluating all covariates at their average value.

\subsection{Regression Results}

Table \ref{very_unfair_logit} shows the estimated marginal effects under different specifications. Column 1 presents the results controlling only for the fixed effects and the Gini coefficient. Column 2 includes basic demographic indicators (age, gender, and marital status). Column 3 includes dummies for maximum educational attainment (the omitted category is completing up to elementary school). Column 4 includes dummies for labor force participation and unemployment. Column 5 includes an index for access to basic services and asset ownership. Column 6 includes political and religious views.

Consistent with the evidence shown in Section \ref{sec_ineq_fairness}, the Gini is positively and statistically significantly related to unfairness perceptions. In a country with average characteristics, a one point decrease in the Gini (from 0.49 to 0.48) decreases the share of the population that believes that the income distribution is very unfair by about half a percentage point ($p < 0.01$). This magnitude does not vary much across specifications (columns 1--6).\footnote{It is important to stress that the interpretation is not necessarily causal. The relationship between income inequality and unfairness perceptions can go both ways. On the one hand, higher inequality can increase the share of the population that believes the distribution is very unfair. But as more people perceive inequality as very unfair, the income distribution can change through several channels (e.g., more corruption or social unrest).}

Several individual-level characteristics predict fairness views. Older people are more likely to perceive the income distribution as very unfair, although the relationship between age and unfairness perception is non-linear. On average, males are just as likely as females to perceive the income distribution as very unfair, while married individuals are slightly less likely to do so. Completing high school is negatively associated with perceptions of unfairness, although the magnitude of the coefficient is small. Being economically active does not seem to be correlated with unfairness views, but being unemployed does. On average, unemployed individuals are about two percentage points more likely to perceive the income distribution as unfair than the employed population. The coefficient on the assets index is negative, suggesting that relatively better-off individuals are less likely to view the income distribution as very unfair, although the coefficient is not statistically different from zero. Ideologically conservative people are statistically less likely to perceive the income distribution as very unfair, although the effect size is small (below half a percentage point). Finally, Catholics are less likely to perceive the income distribution as very unfair.

\subsection{Robustness}

\begin{figure}[htpb]
	\caption{Fairness views and income inequality in Latin America} \label{fig-fairness-gini}
	\centering
	\begin{subfigure}[t]{0.48\textwidth}
		\caption*{Panel A. Correlation between unfairness views and Gini}\label{fig-binscatter-gini-unfair}
		\centering
		\includegraphics[width=\linewidth]{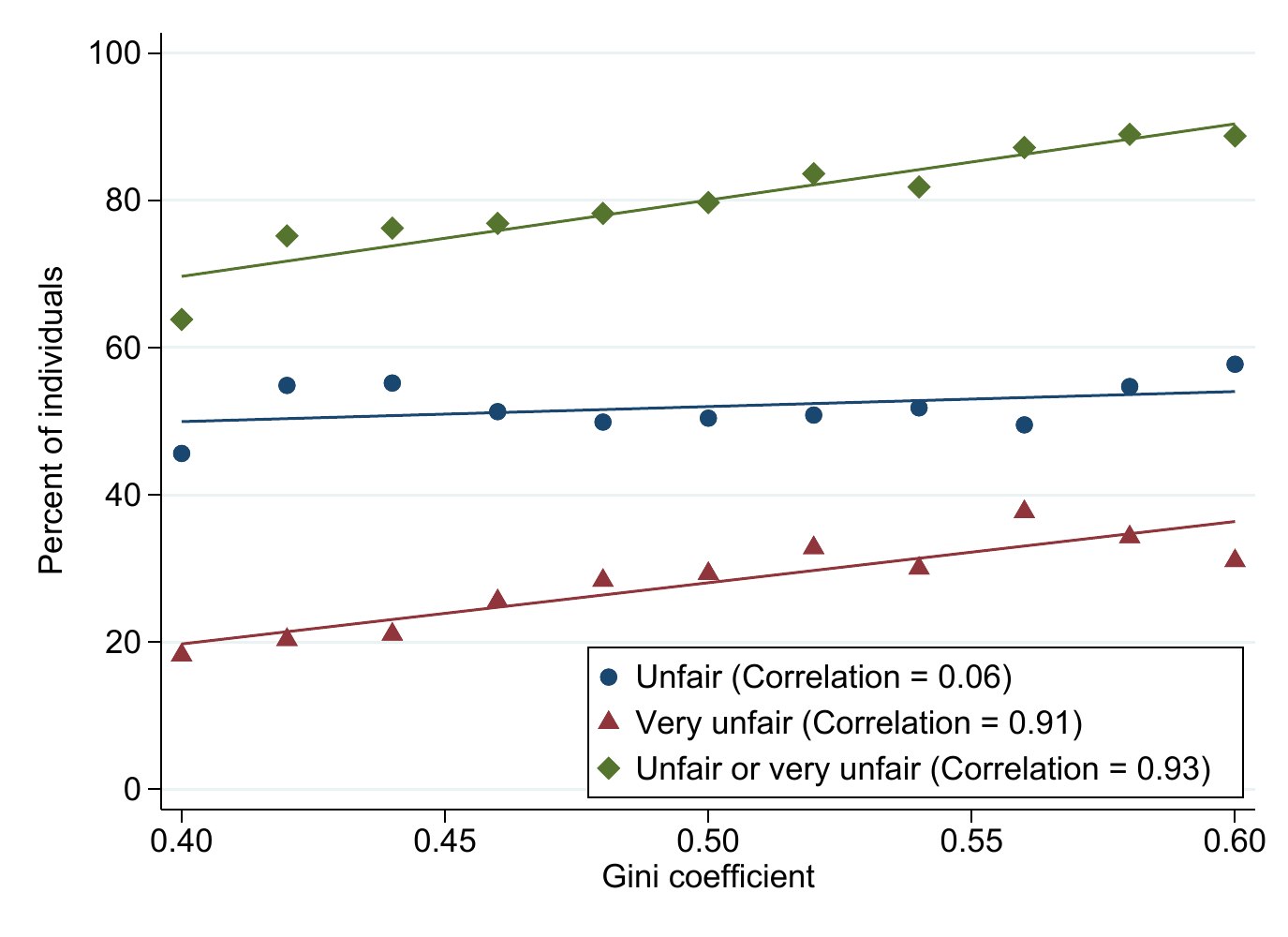}
	\end{subfigure}
	\hfill		
	\begin{subfigure}[t]{0.48\textwidth}
		\caption*{Panel B. Change in fairness and Gini across countries} \label{fig-chg-gini-unfair-0213}
		\includegraphics[width=\linewidth]{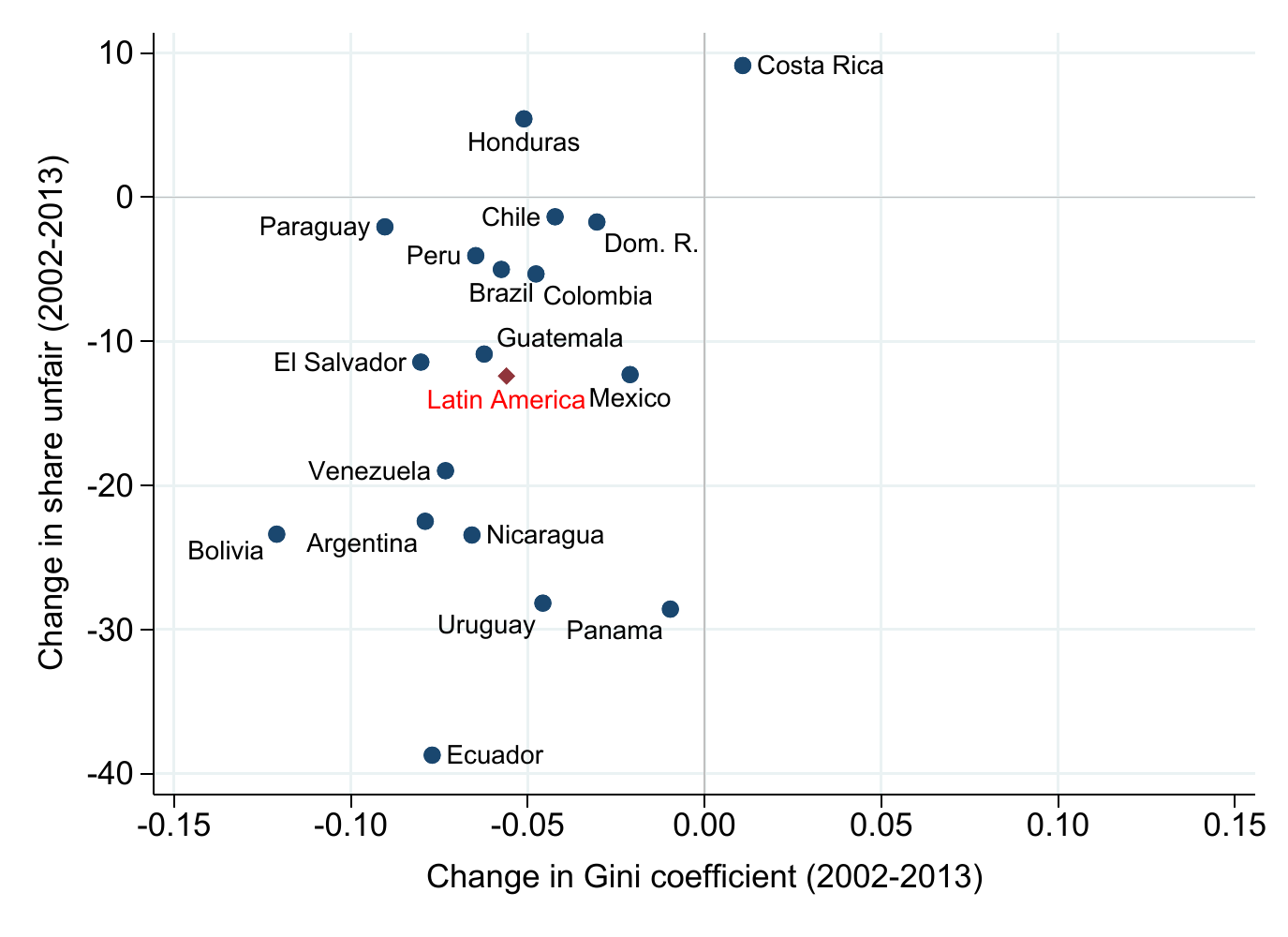}
	\end{subfigure}	
	\hfill				
	{\footnotesize
		\singlespacing \justify
		
		\textbf{Notes:} Panel A shows a binned scatterplot of fairness views and the Gini coefficient. To construct this figure, we group the Ginis of all country-years in bins of width equal to 0.02 Gini points and calculate the average fairness perceptions in each bin.
		
		Panel B plots the percentage point change between 2002 and 2013 in the share of the population that believes that the income distribution is either unfair or very unfair ($y$-axis) against the change in the Gini coefficient between 2002 and 2013 ($x$-axis) for countries in our sample. Due to a break in data comparability or household data unavailability, for some countries, we use inequality data from adjacent years. In 2002, we use: Argentina 2004, Chile 2003, Costa Rica 2010, Ecuador 2003, Guatemala 2006, Nicaragua 2001, Panama 2008, and Peru 2004. In 2013 we use: Guatemala 2014, Mexico 2014, Nicaragua 2014, and Venezuela 2012. See Appendix \ref{sec_data} for more details.
		
	}	
\end{figure}

We conduct three robustness checks. First, we use a broader measure of unfairness that equals one if an individual perceives the income distribution as unfair \textit{or} very unfair as the dependent variable in equation \eqref{eq_fairness}. Appendix Table \ref{unfair_logit} shows the results. The magnitude of the coefficient on the Gini is smaller relative to our baseline specification. Given the relatively large standard errors, the 95\% confidence interval on the Gini includes zero; but the interval also contains the estimated marginal effects in our baseline set of regressions (Table \ref{very_unfair_logit}). In other words, when we use the broader measure of unfairness, our estimates become less informative. This is because strong unfairness views are more strongly correlated with income inequality than weak unfairness views (Table \ref{tab-rel-fairness} and Figure \ref{fig-fairness-gini}).\footnote{The coefficients of some individual-level characteristics are also different when using the broader definition of unfairness views. The effect of completing college on perceptions of unfairness becomes strong and statistically significant. Civil status stops being statistically significant, while the male dummy becomes negative and statistically significant (in both cases consistently so across specifications). The coefficient on the assets index becomes larger and statistically different from zero. Finally, the effect of political ideology and religious views becomes statistically indistinguishable from zero. These results suggest that the population that perceives the income distribution as very unfair tends to be different in observable variables than the population that believes that the income distribution is merely unfair.}

As a second robustness check, we estimate an analogous specification to the one in column 6 of Table \ref{very_unfair_logit}, but controlling for inequality indicators other than the Gini coefficient. Appendix Table \ref{unfair_ineq_logit} shows the results. We find a positive correlation between income inequality and unfairness perceptions across all relative measures of inequality (columns 1--4). The Gini calculated without households with zero income, the Atkinson index, the Theil index, and the Generalized Entropy indicator are consistently correlated with unfairness, and all the coefficients are statistically different from zero at the usual levels. In contrast, the absolute Gini (the only absolute inequality indicator in the table) is negatively correlated with unfairness perceptions, although the coefficient is statistically indistinguishable from zero.

As a final robustness check, we estimate equation \eqref{eq_fairness} using a linear probability model (LPM) instead of a Logit. The choice of a LPM is consistent with the visual evidence shown in Figure \ref{fig-fairness-gini}, Panel A, where fairness views seem to be linearly related to the Gini coefficient. Appendix Table \ref{very_unfair_lpm} shows the results. The estimates are quite similar across specifications. For example, in the specification with the larger set of controls (column 6), the marginal effect of the Gini coefficient is 0.68 in the Logit model and 0.63 in the LPM.

\subsection{Decomposing Changes in Fairness Views Over Time}

\begin{figure}[htp]
	\caption{Oaxaca-Blinder decomposition of unfairness perceptions, 2002-2013}\label{fig-oaxaca}  \centering
	\centering
	\includegraphics[width=.75\linewidth]{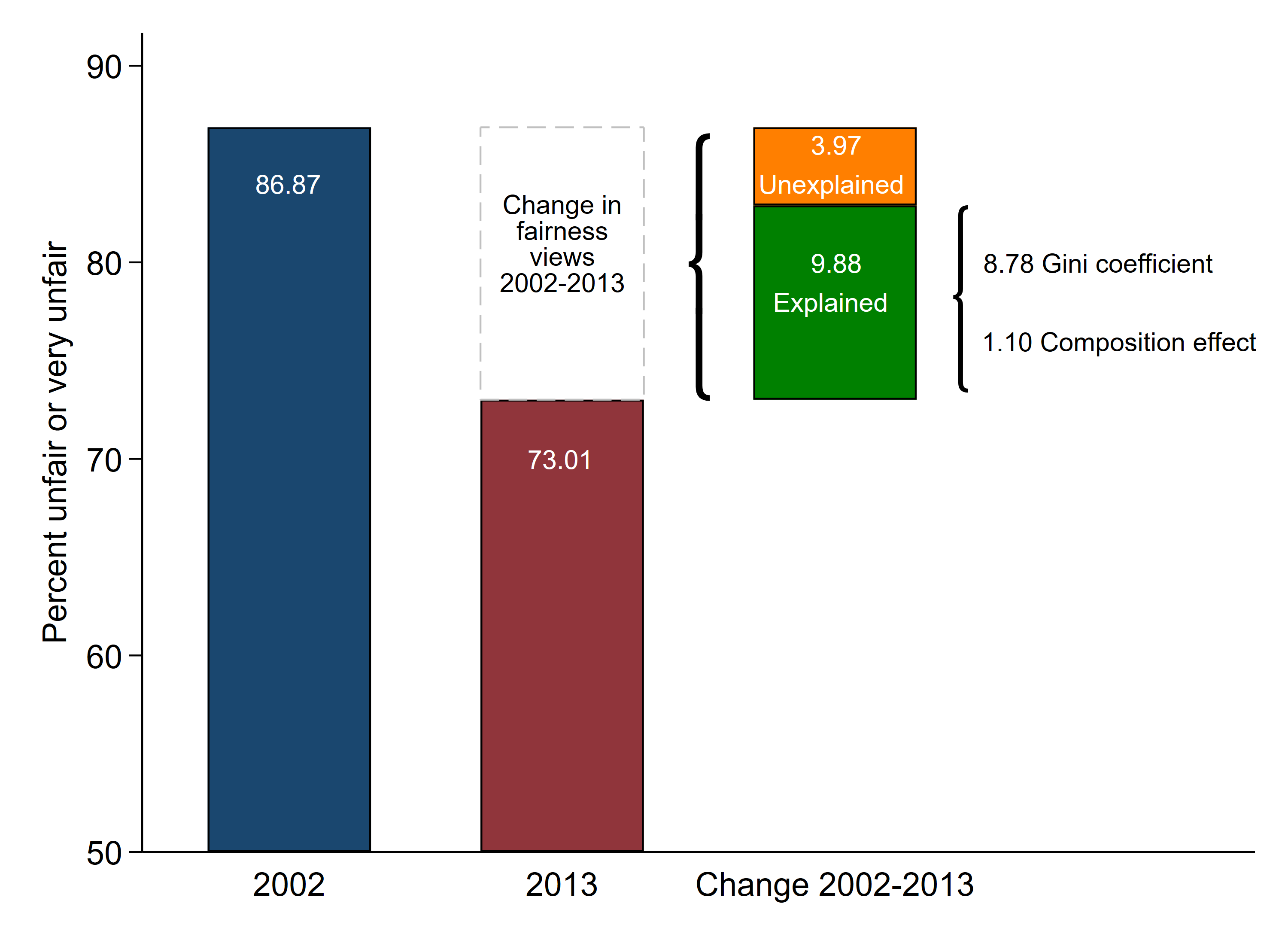}
	\hfill				
	{\footnotesize
		\singlespacing \justify
		
		\textbf{Notes:} This figure presents estimates of the Oaxaca-Blinder decomposition (see Appendix \ref{sec_oaxaca}). The dependent variable is an indicator that equals one for individuals who believe that the income distribution is unfair or very unfair. The regressors in the decomposition include the Gini coefficient, age, age squared, and dummy variables for  marital status, gender, educational attainment, labor force participation, unemployment status, an assets index, political ideology, and religious views. The ``explained'' part of the results refers to changes in the value of the covariables, while the ``unexplained'' refers to changes in the coefficients and the interaction terms.
		
	}	
\end{figure}

Both income inequality and individual-level characteristics are associated with fairness perceptions. Next, we ask which of these two factors mainly explain (in an accounting sense) the reduction in unfairness beliefs over the 2000s. To do this, we perform a Oaxaca-Blinder decomposition, taking 2002 and 2013 as the two groups to be compared (see Appendix \ref{sec_oaxaca} for details on the Oaxaca-Blinder decomposition). In the decomposition, we use the broad definition of unfairness perceptions as the dependent variable and include controls for demographics, educational attainment, employment status, assets, political views, and religion. Figure \ref{fig-oaxaca} summarizes the results.

During 2002--2013, the share of the population perceiving the distribution as unfair decreased 14 percentage points, from 87\% to 73\%. The decomposition suggests that about 28\% of this change (4 percentage points) is accounted for by changes in the elasticity of fairness views to each covariable (i.e., changes in the values of the coefficients in the regression), while the other 72\% can be explained by changes in the covariables' values. Among the covariables included in the decomposition, the one that mainly explains the decline in unfairness perceptions is the change in the Gini, which accounts for 88.9\% of the explained component. In contrast, changes in the composition of the population only account for 11.1\% of the explained component. This result suggests that the decline in unfairness views during the 2000s in Latin America was mainly driven by changes in income inequality and not by changes in the composition of the population.

	\section{The Predictive Power of Fairness Views for Social Unrest} \label{sec_unrest}

There is a vast literature that relates economic inequality---and more recently, measures of polarization---to social cohesion, conflict, and activism.\footnote{For instance, \cite{gasparini2008income} find a strong empirical correlation between inequality and conflict in Latin America. Most previous studies linking inequality and conflict are based on cross-country regressions, and therefore have a notably smaller sample size than our paper.} Arguably, the relationship between income inequality and conflict is partly mediated by fairness views. That is, many individuals mobilize in part because they believe existing inequities are unfair. However, a given level of income inequality might not be seen as unfair by some individuals due to, for example, misperceptions of the actual level of inequality \citep{gimpelson2018misperceiving} or a perception that income gaps are mainly driven by differences in effort \citep{alesina2001doesn}. For these reasons, a regression that links social unrest to income inequality can contain substantial measurement error. We sidetrack these issues by directly measuring the link between social unrest and fairness views.

We measure propensity to engage in social unrest using the opinion polls data. For several political activities, Latinobar\'ometro asks respondents whether they (i) have ever done the activity; (ii) would do the activity; or (iii) would never do the activity. We investigate eight different types of demonstrations: making a complaint on social media, making a complaint to the media, signing a petition, protesting with authorization, protesting without authorization, refusing to pay taxes, participating in riots, and occupying land, factories or buildings. We also create a composite index of political participation which takes the value one if the individual engaged in tax evasion, an illegal protest, signed a petition, or complained to the media, and zero otherwise.\footnote{We use those four measures to construct the index because we do not have data on the other political activities during 2015. Other years have data on fewer political activities.} Participation in these activities is self-reported. Given that respondents had no financial incentives for truth-telling, our results should be taken with caution.

For each activity (and the index), we consider two measures of social unrest. First, we use an indicator that takes the value one if an individual says she has done the activity in the past and zero otherwise. Second, we use an indicator that takes the value one if the individual did the activity in the past \textit{or} says that she is willing to do the activity. We use these measures as dependent variables in Logit regressions. The regressions control for unfairness perceptions, the Gini, and individual-level covariates. Unfortunately, participation in political activities is available only in a few years, so our sample size for these regressions is substantially smaller.

Table \ref{very_unfair_unrest_past}, Panel A shows that unfairness perceptions correlate with participating in political activities in the past. We find positive and statistically significant effects for complaining on social media (column 1) and signing a petition (column 3). Conditional on income inequality, individuals who perceive the income distribution as very unfair are 1.6 percentage points more likely to have complained through social media in the past (from a baseline of 7.8\%) and 1.3 percentage points more likely to have signed a petition in the past (from a baseline of 18.6\%). The rest of the effects tend to be positive, although not statistically different from zero. Conditional on fairness views, we find a statistically significant correlation between the Gini and complaining on social media (column 1), taking part in an authorized demonstration (column 5), and the composite index (column 9). The effect of income inequality on the rest of the activities is statistically indistinguishable from zero. 

Table \ref{very_unfair_unrest_past}, Panel B shows the results when the dependent variable also includes the willingness to participate in the political activities. The set of political activities predicted by fairness views are somewhat different than in Panel A. In Panel B, we find positive and statistically significant effects for complaining through media (either social media or traditional media, columns 1 and 2, respectively) and refusing to pay taxes (column 6). The magnitude of the coefficients that are statistically significant tends to be larger than in the baseline specification. For example, the effect of unfairness views on the propensity to complain on social media is twice as large in Panel B than in Panel A (3.3 vs. 1.6 percentage points, correspondingly). Finally, we find that---holding fairness views constant---income inequality is predictive of refusing to pay taxes (column 4). The effect of income inequality on the rest of the political activities is not statistically different from zero. 

Taken together, these results show that there are political activities for which fairness views and income inequality have predictive power independent of each other (like complaining through social media). However, there are also activities that are exclusively predicted by income inequality (like participating in an unauthorized protest) or fairness views (like signing a petition). This suggests that both fairness views and income inequality capture different channels through which changes in the income distribution can affect social unrest.

	\section{Conclusions} \label{sec_conclusions}

In this paper, we analyze perceptions of distributive justice in a context of falling income inequality. We show that fairness beliefs moved in line with the evolution of objective inequality indicators: both unfairness perceptions and income inequality declined across countries and over time in our sample. Some individual-level characteristics, such as employment status and political ideology, are systematically correlated with fairness views. Fairness views have predictive power for self-reported propensity to mobilize above and beyond income inequality (and vice-versa).

Our findings are relevant for both researchers and policymakers. For researchers, our results suggest that, in some contexts, one can proxy fairness views using relative measures of income inequality, such as the widely used Gini coefficient. This is reassuring since inequality measures are much more widely available than fairness views in standard datasets. 

For policymakers, our findings warn about concerning levels of dissatisfaction with existing income disparities. Three in four Latin Americans believe that the income distribution is unfair, and such perceptions have proved to be relatively inelastic to a large compression of the income distribution by historical standards. If fairness perceptions are interpreted as preferences for some leveling of income, our results indicate that a striking majority is in favor of reducing the existing disparities between the rich and the poor, while relatively few people believe that income disparities should remain the same. A second actionable insight for policymakers is that fairness views act as a thermometer of individuals' latent propensity to engage in political activities. Thus, if policymakers want to prevent social unrest, they ought to pay attention to the evolution of fairness views and take preventive measures before the majority of people perceive the income distribution as unfair.

A caveat with our results is that we cannot tell whether most individuals believe that the income distribution is unfair because (i) they have inaccurate views about the level of income inequality (perhaps, believing that income is more unequally distributed than it objectively is); or (ii) individuals accurately assess the level of inequality and believe that existing inequities are unfair (perhaps, because the process that generates income differences is not fair or because existing inequities are too large). Disentangling the contribution of these and other channels is a challenge for future research.

	\clearpage 
	\section*{Tables and Figures}

\begin{table}[htbp]{\footnotesize
		\begin{center}
			\caption{Correlation between inequality indicators and fairness views, 1997-2015}\label{tab-rel-fairness}
			\begin{tabular}{lccccccccccc}
				\midrule
				& \multicolumn{3}{c}{} &   & \multicolumn{3}{c}{} &   & \multicolumn{3}{c}{Averaging correlations} \\
				& \multicolumn{3}{c}{Individual-level data} &   & \multicolumn{3}{c}{Country-by-year level data} &   & \multicolumn{3}{c}{across countries} \\
				\cmidrule{2-4}\cmidrule{6-8}\cmidrule{10-12}      &  U./V.U.  & V.U. & U. &   &  U./V.U.  & V.U. & U. &   &  U./V.U.  & V.U. & U. \\
				&  (1)  & (2) & (3) &   &  (4)  & (5) & (6) &   &  (7)  & (8) & (9) \\
				\midrule
				Gini coefficient & 0.39 & 0.36 & 0.10 &   & 0.84 & 0.82 & 0.25 &   & 0.39 & 0.28 & 0.15 \\
				& (0.07) & (0.07) & (0.09) &   & (0.10) & (0.16) & (0.35) &   & (0.07) & (0.07) & (0.09) \\
				Theil index & 0.39 & 0.38 & 0.05 &   & 0.85 & 0.82 & 0.27 &   & 0.33 & 0.20 & 0.18 \\
				& (0.07) & (0.07) & (0.09) &   & (0.09) & (0.16) & (0.35) &   & (0.07) & (0.07) & (0.09) \\
				Atkinson, A(0.5) & 0.38 & 0.35 & 0.10 &   & 0.84 & 0.82 & 0.25 &   & 0.37 & 0.26 & 0.15 \\
				& (0.07) & (0.07) & (0.09) &   & (0.10) & (0.16) & (0.35) &   & (0.07) & (0.07) & (0.09) \\
				Atkinson, A(1) & 0.36 & 0.30 & 0.13 &   & 0.84 & 0.82 & 0.25 &   & 0.38 & 0.28 & 0.13 \\
				& (0.07) & (0.08) & (0.09) &   & (0.10) & (0.16) & (0.34) &   & (0.07) & (0.08) & (0.09) \\
				Mean log deviation & 0.35 & 0.29 & 0.14 &   & 0.84 & 0.82 & 0.25 &   & 0.38 & 0.28 & 0.13 \\
				& (0.07) & (0.08) & (0.09) &   & (0.10) & (0.16) & (0.34) &   & (0.07) & (0.08) & (0.09) \\
				Coefficient Variation & 0.33 & 0.36 & 0.00 &   & 0.78 & 0.78 & 0.19 &   & 0.18 & 0.20 & 0.10 \\
				& (0.08) & (0.08) & (0.09) &   & (0.13) & (0.16) & (0.36) &   & (0.08) & (0.08) & (0.09) \\
				Ratio 75/25 & 0.29 & 0.15 & 0.24 &   & 0.80 & 0.78 & 0.26 &   & 0.36 & 0.29 & 0.12 \\
				& (0.07) & (0.09) & (0.08) &   & (0.11) & (0.17) & (0.33) &   & (0.07) & (0.09) & (0.08) \\
				Generalized entropy & 0.29 & 0.35 & -0.04 &   & 0.80 & 0.72 & 0.35 &   & 0.18 & 0.09 & 0.19 \\
				& (0.05) & (0.08) & (0.08) &   & (0.11) & (0.18) & (0.34) &   & (0.05) & (0.08) & (0.08) \\
				Ratio 90/10 & 0.23 & 0.10 & 0.21 &   & 0.81 & 0.79 & 0.25 &   & 0.30 & 0.30 & 0.07 \\
				& (0.07) & (0.08) & (0.08) &   & (0.11) & (0.17) & (0.32) &   & (0.07) & (0.08) & (0.08) \\
				Variance & -0.08 & -0.01 & -0.12 &   & -0.25 & 0.10 & -0.71 &   & -0.06 & 0.04 & -0.12 \\
				& (0.07) & (0.08) & (0.08) &   & (0.39) & (0.43) & (0.14) &   & (0.07) & (0.08) & (0.08) \\
				Absolute Gini & -0.21 & -0.10 & -0.18 &   & -0.71 & -0.46 & -0.64 &   & -0.18 & -0.10 & -0.22 \\
				& (0.09) & (0.10) & (0.08) &   & (0.23) & (0.26) & (0.31) &   & (0.09) & (0.10) & (0.08) \\
				Kolm, K(1) & -0.31 & -0.18 & -0.22 &   & -0.80 & -0.64 & -0.50 &   & -0.22 & -0.16 & -0.22 \\
				& (0.09) & (0.10) & (0.08) &   & (0.12) & (0.18) & (0.37) &   & (0.09) & (0.10) & (0.08) \\
				\midrule
			\end{tabular}
		\end{center}
		\begin{singlespace}  \vspace{-.5cm}
			\noindent \justify \textbf{Notes:} This table presents correlations between fairness views and income inequality. In columns 1--3, we calculate the correlations at the individual-level pooling all countries and years in our sample. In columns 4--6, we calculate the average unfairness views in each country-year and then calculate the correlation between each inequality indicator in the corresponding country-year and the average fairness views. In columns 7--9, we calculate the correlation between each inequality indicator and fairness views over time for each country separately (using the individual-level data) and then average the correlations across countries. U./V.U. stands for ``Unfair or Very Unfair''; V.U. stands for ``Very Unfair''; and U. stands for ``Unfair.'' Boostrapped standard errors are in parenthesis. 
		\end{singlespace} 	
	}
\end{table}

\clearpage
\begin{table}[htpb!]{\footnotesize
		\begin{center}
			\caption{Logit regressions of unfairness perceptions (very unfair) and individuals' characteristics} \label{very_unfair_logit}
			\newcommand\w{1.30}
			\begin{tabular}{l@{}lR{\w cm}@{}L{0.43cm}R{\w cm}@{}L{0.43cm}R{\w cm}@{}L{0.43cm}R{\w cm}@{}L{0.43cm}R{\w cm}@{}L{0.43cm}R{\w cm}@{}L{0.43cm}}
				\midrule
				&& 	\multicolumn{12}{c}{Dependent Variable: Believes income distribution is very unfair}  \\\cmidrule{3-14} 
				&& (1) && (2) && (3) && (4) && (5) && (6) \\	
				\midrule 
				Gini&&0.679&\sym{***}&0.674&\sym{***}&0.678&\sym{***}&0.680&\sym{***}&0.678&\sym{***}&0.684&\sym{***}\\
&&(0.227&)&(0.225&)&(0.225&)&(0.225&)&(0.225&)&(0.223&)\\
Age&&&&0.004&\sym{***}&0.004&\sym{***}&0.004&\sym{***}&0.004&\sym{***}&0.004&\sym{***}\\
&&&&(0.000&)&(0.000&)&(0.000&)&(0.000&)&(0.000&)\\
Age squared&&&&$-$0.000&\sym{***}&$-$0.000&\sym{***}&$-$0.000&\sym{***}&$-$0.000&\sym{***}&$-$0.000&\sym{***}\\
&&&&(0.000&)&(0.000&)&(0.000&)&(0.000&)&(0.000&)\\
Male&&&&$-$0.002&&$-$0.002&&$-$0.001&&$-$0.001&&0.001&\\
&&&&(0.003&)&(0.003&)&(0.003&)&(0.003&)&(0.003&)\\
Married&&&&$-$0.006&\sym{*}&$-$0.006&\sym{**}&$-$0.006&\sym{*}&$-$0.006&\sym{*}&$-$0.005&\\
&&&&(0.003&)&(0.003&)&(0.003&)&(0.003&)&(0.003&)\\
Finished HS&&&&&&$-$0.010&\sym{*}&$-$0.009&\sym{*}&$-$0.008&\sym{*}&$-$0.006&\\
&&&&&&(0.005&)&(0.005&)&(0.005&)&(0.005&)\\
Finished coll.&&&&&&$-$0.001&&0.000&&0.002&&0.006&\\
&&&&&&(0.007&)&(0.007&)&(0.006&)&(0.006&)\\
In labor force&&&&&&&&$-$0.003&&$-$0.003&&$-$0.002&\\
&&&&&&&&(0.004&)&(0.004&)&(0.003&)\\
Unemployed&&&&&&&&0.020&\sym{***}&0.019&\sym{***}&0.019&\sym{***}\\
&&&&&&&&(0.007&)&(0.007&)&(0.007&)\\
Assets index&&&&&&&&&&$-$0.009&&$-$0.007&\\
&&&&&&&&&&(0.006&)&(0.006&)\\
Conservative&&&&&&&&&&&&$-$0.003&\sym{*}\\
&&&&&&&&&&&&(0.002&)\\
Catholic&&&&&&&&&&&&$-$0.011&\sym{***}\\
&&&&&&&&&&&&(0.004&)\\
N&&167,436&&166,105&&164,662&&164,436&&164,436&&164,436&\\

				\midrule
			\end{tabular}
		\end{center}
		\begin{singlespace}  \vspace{-.5cm}
			\noindent \justify \textbf{Notes:} This table shows estimates of the relationship between an indicator that equals one for individuals who believe that the income distribution is very unfair and the Gini coefficient controlling for individuals' characteristics. Coefficients are estimated through Logit regressions and represent the marginal effects evaluated at the mean values of the rest of the variables. Observations are weighted by the individual's probability of being interviewed. All specifications include country and year fixed effects. $^{***}$, $^{**}$ and $^*$ denote significance at 10\%, 5\% and 1\% levels, respectively. Heteroskedasticity-robust standard errors clustered at the country-by-year level in parentheses. 
		\end{singlespace} 	
	}
\end{table}

\begin{landscape}
	
	\begin{table}[htpb!]{\footnotesize
			\begin{center}
				\caption{Logit regressions of unfairness perceptions (very unfair) and activism}\label{very_unfair_unrest_past}
				\newcommand\w{1.58}
				\begin{tabular}{l@{}lR{\w cm}@{}L{0.43cm}R{\w cm}@{}L{0.43cm}R{\w cm}@{}L{0.43cm}R{\w cm}@{}L{0.43cm}R{\w cm}@{}L{0.43cm}R{\w cm}@{}L{0.43cm}R{\w cm}@{}L{0.43cm}R{\w cm}@{}L{0.43cm}R{\w cm}@{}L{0.43cm}}
					\midrule
					&& Complain   &&            &&          &&             &&          && Refuse  &&             && Occupy    && Activism \\
					&& on social  && Complain  && Sign a    && Authorized  && Unauth.  &&  to pay && Participate && land or   && composite \\
					&& media      && to media  && petition  && protest     && protest  && taxes   && in riots    && buildings && index \\
					&& (1)        &&     (2)   && (3)       &&  (4)        && (5)      && (6)     && (7)        && (8)        && (9) \\
					\midrule

									\addlinespace
					\multicolumn{14}{l}{\hspace{-1em} \textbf{Panel A.\ Dependent variable: Have done the activity in the past}} \\
					\midrule

					Very unfair&&0.016&\sym{***}&0.005&&0.013&\sym{**}&0.011&&$-$0.001&&0.004&&0.002&&0.003&&0.006&\\
&&(0.004&)&(0.004&)&(0.006&)&(0.007&)&(0.004&)&(0.005&)&(0.002&)&(0.002&)&(0.013&)\\
Gini&&0.322&\sym{*}&0.209&&$-$0.123&&$-$0.060&&0.164&\sym{*}&0.105&&0.015&&$-$0.054&&0.643&\sym{*}\\
&&(0.175&)&(0.182&)&(0.287&)&(0.149&)&(0.092&)&(0.089&)&(0.034&)&(0.040&)&(0.336&)\\
Mean Dep. Var.&&0.078&&0.072&&0.186&&0.080&&0.046&&0.043&&0.011&&0.015&&0.251&\\

					\midrule
					
					\addlinespace
					\multicolumn{14}{l}{\hspace{-1em} \textbf{Panel B.\ Dependent variable: Have done the activity in the past or would do the activity}} \\ 
					\midrule
					Very unfair&&0.033&\sym{**}&0.031&\sym{*}&$-$0.012&&0.025&&0.016&&0.024&\sym{**}&$-$0.003&&$-$0.001&&0.011&\\
&&(0.015&)&(0.017&)&(0.008&)&(0.025&)&(0.014&)&(0.010&)&(0.006&)&(0.008&)&(0.015&)\\
Gini&&0.627&&0.603&&$-$0.597&&0.162&&0.452&&0.629&\sym{*}&0.039&&0.014&&0.954&\\
&&(0.786&)&(0.842&)&(0.479&)&(0.661&)&(0.446&)&(0.328&)&(0.107&)&(0.147&)&(0.695&)\\
Mean Dep. Var.&&0.413&&0.477&&0.527&&0.296&&0.210&&0.187&&0.063&&0.084&&0.688&\\

					\midrule
					
					\addlinespace
					N&&18,605&&18,827&&52,704&&17,475&&18,846&&18,553&&16,823&&16,779&&18,155&\\

					\midrule

				\end{tabular}
			\end{center}
			\begin{singlespace} 
				\noindent \justify \textbf{Notes:} This table shows estimates of the determinants of  participating in the political activity listed in the column header. Coefficients are estimated through Logit regressions and represent the marginal effects evaluated at the mean values of the rest of the variables. Observations are weighted by the individual's probability of being interviewed. All specifications control for age, age squared, gender, marital status, maximum educational attainment, labor force participation, unemployment status, assets index, political ideology, and religion. The regression in column 3 also controls for country and year fixed effects. Column 9 shows a composite index which takes the value one if the individual reports having engaged in tax evasion, an illegal protest, signed a petition, or complained in the media, and zero otherwise. The rest of the political activities are available in only one year and thus we cannot include fixed effects. Participation in political activities is self-reported.  $^{***}$, $^{**}$ and $^*$ denote significance at 10\%, 5\% and 1\% levels, respectively. Heteroskedasticity-robust standard errors clustered at the country-by-year level in parentheses. 
			\end{singlespace} 	
		}
	\end{table}

\end{landscape}

\clearpage

\clearpage 

\clearpage 

	\clearpage		
	\begin{singlespace}
	 \bibliographystyle{chicago}
	 \bibliography{0-fairness}                                                                                     

\begin{thebibliography}{}

\bibitem[\protect\citeauthoryear{Alesina and Giuliano}{Alesina and
  Giuliano}{2011}]{alesina2011preferences}
Alesina, A. and P.~Giuliano (2011).
\newblock Preferences for redistribution.
\newblock In {\em Handbook of social economics}, Volume~1, pp.\  93--131.
  Elsevier.

\bibitem[\protect\citeauthoryear{Alesina, Glaeser, and Sacerdote}{Alesina
  et~al.}{2001}]{alesina2001doesn}
Alesina, A., E.~Glaeser, and B.~Sacerdote (2001).
\newblock Why doesn't the us have a european-style welfare system?
\newblock Technical report, National bureau of economic research.

\bibitem[\protect\citeauthoryear{Alesina and Perotti}{Alesina and
  Perotti}{1996}]{alesina1996income}
Alesina, A. and R.~Perotti (1996).
\newblock Income distribution, political instability, and investment.
\newblock {\em European economic review\/}~{\em 40\/}(6), 1203--1228.

\bibitem[\protect\citeauthoryear{Alesina, Stantcheva, and Teso}{Alesina
  et~al.}{2018}]{alesina2018intergenerational}
Alesina, A., S.~Stantcheva, and E.~Teso (2018).
\newblock Intergenerational mobility and preferences for redistribution.
\newblock {\em American Economic Review\/}~{\em 108\/}(2), 521--54.

\bibitem[\protect\citeauthoryear{Alm{\aa}s, Cappelen, and Tungodden}{Alm{\aa}s
  et~al.}{2020}]{almaas2020cutthroat}
Alm{\aa}s, I., A.~W. Cappelen, and B.~Tungodden (2020).
\newblock Cutthroat capitalism versus cuddly socialism: Are americans more
  meritocratic and efficiency-seeking than scandinavians?
\newblock {\em Journal of Political Economy\/}~{\em 128\/}(5), 000--000.

\bibitem[\protect\citeauthoryear{Alvaredo and Gasparini}{Alvaredo and
  Gasparini}{2015}]{alvaredo2015recent}
Alvaredo, F. and L.~Gasparini (2015).
\newblock Recent trends in inequality and poverty in developing countries.
\newblock {\em Handbook of Income Distribution\/}~{\em 2}, 697--805.

\bibitem[\protect\citeauthoryear{Amiel and Cowell}{Amiel and
  Cowell}{1999}]{amiel1999thinking}
Amiel, Y. and F.~Cowell (1999).
\newblock {\em Thinking about inequality: Personal judgment and income
  distributions}.
\newblock Cambridge University Press.

\bibitem[\protect\citeauthoryear{Amiel and Cowell}{Amiel and
  Cowell}{1992}]{amiel1992measurement}
Amiel, Y. and F.~A. Cowell (1992).
\newblock Measurement of income inequality: Experimental test by questionnaire.
\newblock {\em Journal of Public Economics\/}~{\em 47\/}(1), 3--26.

\bibitem[\protect\citeauthoryear{Atkinson and Brandolini}{Atkinson and
  Brandolini}{2010}]{atkinson2010analyzing}
Atkinson, A.~B. and A.~Brandolini (2010).
\newblock On analyzing the world distribution of income.
\newblock {\em The World Bank Economic Review\/}~{\em 24\/}(1), 1--37.

\bibitem[\protect\citeauthoryear{Atkinson, Piketty, and Saez}{Atkinson
  et~al.}{2011}]{atkinson2011top}
Atkinson, A.~B., T.~Piketty, and E.~Saez (2011).
\newblock Top incomes in the long run of history.
\newblock {\em Journal of Economic Literature\/}~{\em 49\/}(1), 3--71.

\bibitem[\protect\citeauthoryear{Cappelen, Hole, and S{\o}rensen}{Cappelen
  et~al.}{2007}]{cappelen2007pluralism}
Cappelen, A.~W., A.~D. Hole, and B.~S{\o}rensen, Erikand~Tungodden (2007).
\newblock The pluralism of fairness ideals: An experimental approach.
\newblock {\em American Economic Review\/}~{\em 97\/}(3), 818--827.

\bibitem[\protect\citeauthoryear{Cappelen, Konow, S{\o}rensen, and
  Tungodden}{Cappelen et~al.}{2013}]{cappelen2013just}
Cappelen, A.~W., J.~Konow, E.~S{\o}rensen, and B.~Tungodden (2013).
\newblock Just luck: An experimental study of risk-taking and fairness.
\newblock {\em American Economic Review\/}~{\em 103\/}(4), 1398--1413.

\bibitem[\protect\citeauthoryear{CEPAL}{CEPAL}{2010}]{cepal2010america}
CEPAL (2010).
\newblock Am{\'e}rica latina frente al espejo: dimensiones objetivas y
  subjetivas de la inequidad social y el bienestar en la regi{\'o}n.

\bibitem[\protect\citeauthoryear{Choi}{Choi}{2019}]{choi2019revisiting}
Choi, G. (2019).
\newblock Revisiting the redistribution hypothesis with perceived inequality
  and redistributive preferences.
\newblock {\em European Journal of Political Economy\/}~{\em 58}, 220--244.

\bibitem[\protect\citeauthoryear{Cozzolino}{Cozzolino}{2011}]{cozzolino2011trust}
Cozzolino, P.~J. (2011).
\newblock Trust, cooperation, and equality: a psychological analysis of the
  formation of social capital.
\newblock {\em British Journal of Social Psychology\/}~{\em 50\/}(2), 302--320.

\bibitem[\protect\citeauthoryear{Cruces, Perez-Truglia, and Tetaz}{Cruces
  et~al.}{2013}]{cruces2013biased}
Cruces, G., R.~Perez-Truglia, and M.~Tetaz (2013).
\newblock Biased perceptions of income distribution and preferences for
  redistribution: Evidence from a survey experiment.
\newblock {\em Journal of Public Economics\/}~{\em 98}, 100--112.

\bibitem[\protect\citeauthoryear{Esteban, Mayoral, and Ray}{Esteban
  et~al.}{2012}]{esteban2012ethnicity}
Esteban, J., L.~Mayoral, and D.~Ray (2012).
\newblock Ethnicity and conflict: An empirical study.
\newblock {\em American Economic Review\/}~{\em 102\/}(4), 1310--42.

\bibitem[\protect\citeauthoryear{Esteban and Ray}{Esteban and
  Ray}{2011}]{esteban2011linking}
Esteban, J. and D.~Ray (2011).
\newblock Linking conflict to inequality and polarization.
\newblock {\em American Economic Review\/}~{\em 101\/}(4), 1345--74.

\bibitem[\protect\citeauthoryear{Fehr, Mollerstrom, and Perez-Truglia}{Fehr
  et~al.}{2021}]{fehr2021your}
Fehr, D., J.~Mollerstrom, and R.~Perez-Truglia (2021).
\newblock Your place in the world-relative income and global inequality.

\bibitem[\protect\citeauthoryear{Fehr, Mueller, and Preuss}{Fehr
  et~al.}{2020}]{preuss2020}
Fehr, D., D.~Mueller, and M.~Preuss (2020).
\newblock Social mobility perceptions and inequality acceptance.
\newblock {\em Working Paper\/}.

\bibitem[\protect\citeauthoryear{Finseraas}{Finseraas}{2009}]{finseraas2009income}
Finseraas, H. (2009).
\newblock Income inequality and demand for redistribution: A multilevel
  analysis of european public opinion.
\newblock {\em Scandinavian Political Studies\/}~{\em 32\/}(1), 94--119.

\bibitem[\protect\citeauthoryear{Gasparini, Cruces, Tornarolli, and
  Mej{\'i}a}{Gasparini et~al.}{2011}]{gasparini2011recent}
Gasparini, L., G.~Cruces, L.~Tornarolli, and D.~Mej{\'i}a (2011).
\newblock Recent trends in income inequality in latin america.
\newblock {\em Economia\/}~{\em 11\/}(2), 147--201.

\bibitem[\protect\citeauthoryear{Gasparini, Horenstein, Molina, and
  Olivieri}{Gasparini et~al.}{2008}]{gasparini2008income}
Gasparini, L., M.~Horenstein, E.~Molina, and S.~Olivieri (2008).
\newblock Income polarization in latin america: Patterns and links with
  institutions and conflict.
\newblock {\em Oxford Development Studies\/}~{\em 36\/}(4), 461--484.

\bibitem[\protect\citeauthoryear{Gasparini and Lustig}{Gasparini and
  Lustig}{2011}]{gasparini2011rise}
Gasparini, L. and N.~Lustig (2011).
\newblock The rise and fall of income inequality in latin america.
\newblock Technical report, Documento de Trabajo.

\bibitem[\protect\citeauthoryear{Gimpelson and Treisman}{Gimpelson and
  Treisman}{2018}]{gimpelson2018misperceiving}
Gimpelson, V. and D.~Treisman (2018).
\newblock Misperceiving inequality.
\newblock {\em Economics \& Politics\/}~{\em 30\/}(1), 27--54.

\bibitem[\protect\citeauthoryear{Gustavsson and Jordahl}{Gustavsson and
  Jordahl}{2008}]{gustavsson2008inequality}
Gustavsson, M. and H.~Jordahl (2008).
\newblock Inequality and trust in sweden: Some inequalities are more harmful
  than others.
\newblock {\em Journal of Public Economics\/}~{\em 92\/}(1-2), 348--365.

\bibitem[\protect\citeauthoryear{Hvidberg, Kreiner, and Stantcheva}{Hvidberg
  et~al.}{2020}]{hvidberg2020social}
Hvidberg, K.~B., C.~Kreiner, and S.~Stantcheva (2020).
\newblock Social position and fairness views.
\newblock Technical report, National Bureau of Economic Research.

\bibitem[\protect\citeauthoryear{Karadja, Mollerstrom, and Seim}{Karadja
  et~al.}{2017}]{karadja2017richer}
Karadja, M., J.~Mollerstrom, and D.~Seim (2017).
\newblock Richer (and holier) than thou? the effect of relative income
  improvements on demand for redistribution.
\newblock {\em Review of Economics and Statistics\/}~{\em 99\/}(2), 201--212.

\bibitem[\protect\citeauthoryear{Kuziemko, Norton, Saez, and
  Stantcheva}{Kuziemko et~al.}{2015}]{kuziemko2015elastic}
Kuziemko, I., M.~I. Norton, E.~Saez, and S.~Stantcheva (2015).
\newblock How elastic are preferences for redistribution? evidence from
  randomized survey experiments.
\newblock {\em American Economic Review\/}~{\em 105\/}(4), 1478--1508.

\bibitem[\protect\citeauthoryear{Lustig, Lopez-Calva, and Ortiz-Juarez}{Lustig
  et~al.}{2013}]{lustig2013declining}
Lustig, N., L.~F. Lopez-Calva, and E.~Ortiz-Juarez (2013).
\newblock Declining inequality in latin america in the 2000s: The cases of
  argentina, brazil, and mexico.
\newblock {\em World development\/}~{\em 44}, 129--141.

\bibitem[\protect\citeauthoryear{Mart{\'\i}nez~Correa, Pe{\~n}aloza~Pacheco,
  and Gasparini}{Mart{\'\i}nez~Correa et~al.}{2020}]{martinez2020latin}
Mart{\'\i}nez~Correa, J., L.~J. Pe{\~n}aloza~Pacheco, and L.~C. Gasparini
  (2020).
\newblock Latin american brotherhood?: immigration and preferences for
  redistribution.
\newblock {\em Documentos de Trabajo del CEDLAS\/}.

\bibitem[\protect\citeauthoryear{Meltzer and Richard}{Meltzer and
  Richard}{1981}]{meltzer1981rational}
Meltzer, A.~H. and S.~F. Richard (1981).
\newblock A rational theory of the size of government.
\newblock {\em Journal of Political Economy\/}~{\em 89\/}(5), 914--927.

\bibitem[\protect\citeauthoryear{Milanovic}{Milanovic}{2016}]{milanovic2016global}
Milanovic, B. (2016).
\newblock {\em Global inequality: A new approach for the age of globalization}.
\newblock Harvard University Press.

\bibitem[\protect\citeauthoryear{Neville}{Neville}{2012}]{neville2012economic}
Neville, L. (2012).
\newblock Do economic equality and generalized trust inhibit academic
  dishonesty? evidence from state-level search-engine queries.
\newblock {\em Psychological Science\/}~{\em 23\/}(4), 339--345.

\bibitem[\protect\citeauthoryear{Norton and Ariely}{Norton and
  Ariely}{2011}]{norton2011building}
Norton, M.~I. and D.~Ariely (2011).
\newblock Building a better america—one wealth quintile at a time.
\newblock {\em Perspectives on psychological science\/}~{\em 6\/}(1), 9--12.

\bibitem[\protect\citeauthoryear{Oishi, Kesebir, and Diener}{Oishi
  et~al.}{2011}]{oishi2011income}
Oishi, S., S.~Kesebir, and E.~Diener (2011).
\newblock Income inequality and happiness.
\newblock {\em Psychological science\/}~{\em 22\/}(9), 1095--1100.

\bibitem[\protect\citeauthoryear{Page and Goldstein}{Page and
  Goldstein}{2016}]{page2016subjective}
Page, L. and D.~G. Goldstein (2016).
\newblock Subjective beliefs about the income distribution and preferences for
  redistribution.
\newblock {\em Social Choice and Welfare\/}~{\em 47\/}(1), 25--61.

\bibitem[\protect\citeauthoryear{Ravallion}{Ravallion}{2003}]{ravallion2003debate}
Ravallion, M. (2003).
\newblock The debate on globalization, poverty and inequality: why measurement
  matters.
\newblock {\em International affairs\/}~{\em 79\/}(4), 739--753.

\bibitem[\protect\citeauthoryear{Rodr{\'\i}guez}{Rodr{\'\i}guez}{2014}]{rodriguez2014percepciones}
Rodr{\'\i}guez, S.~A. (2014).
\newblock Percepciones de desigualdad socioecon{\'o}mica: Un estudio
  exploratorio para el caso argentino.
\newblock {\em Revista de Ciencias Sociales\/}~{\em 27\/}(34), 93--118.

\bibitem[\protect\citeauthoryear{{World Bank}}{{World
  Bank}}{2016}]{world2016poverty}
{World Bank} (2016).
\newblock Poverty and shared prosperity 2016: Taking on inequality.

\bibitem[\protect\citeauthoryear{Zmerli and Castillo}{Zmerli and
  Castillo}{2015}]{zmerli2015income}
Zmerli, S. and J.~C. Castillo (2015).
\newblock Income inequality, distributive fairness and political trust in latin
  america.
\newblock {\em Social science research\/}~{\em 52}, 179--192.

\end{thebibliography}
	\end{singlespace}

	\clearpage 
	\appendix 
	\begin{center}\noindent{\LARGE \textbf{Appendix}}\end{center}

\section{Additional Figures and Tables} \label{app_add_figs}

\setcounter{table}{0}
\setcounter{figure}{0}
\setcounter{equation}{0}	
\renewcommand{\thetable}{A\arabic{table}}
\renewcommand{\thefigure}{A\arabic{figure}}
\renewcommand{\theequation}{A\arabic{equation}}

\begin{figure}[htpb]
	\caption{Perceptions of unfairness and individual characteristics, 1997--2015} \label{fig-fairness-grps}
	\centering
	\begin{subfigure}[t]{.48\textwidth}
		\caption*{Panel A. By age}\label{fig-fairness-age}
		\centering
		\includegraphics[width=\linewidth]{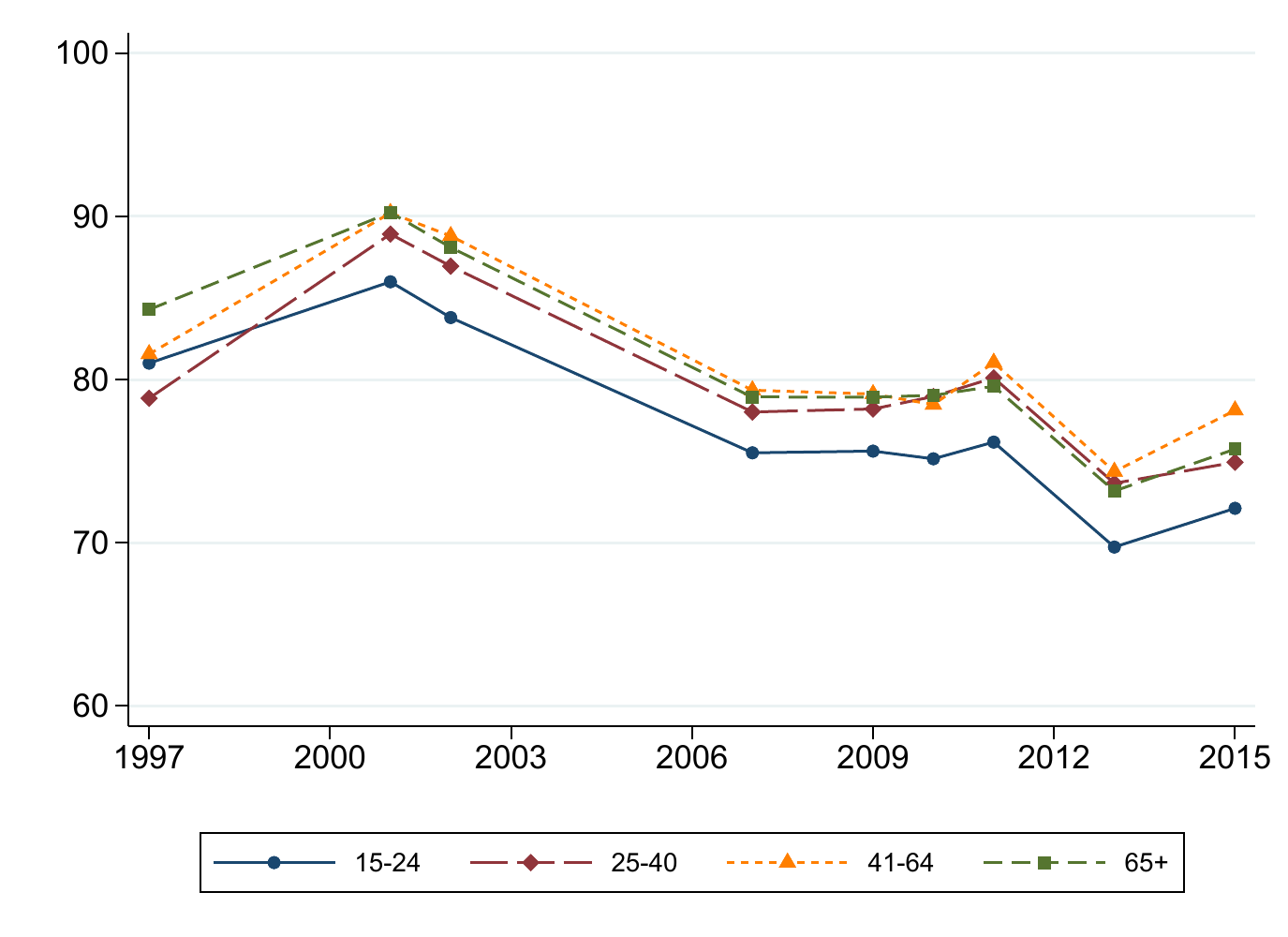}
	\end{subfigure}
	\hfill		
	\begin{subfigure}[t]{0.48\textwidth}
		\caption*{Panel B. By sex}\label{fig-fairness-sex}
		\centering
		\includegraphics[width=\linewidth]{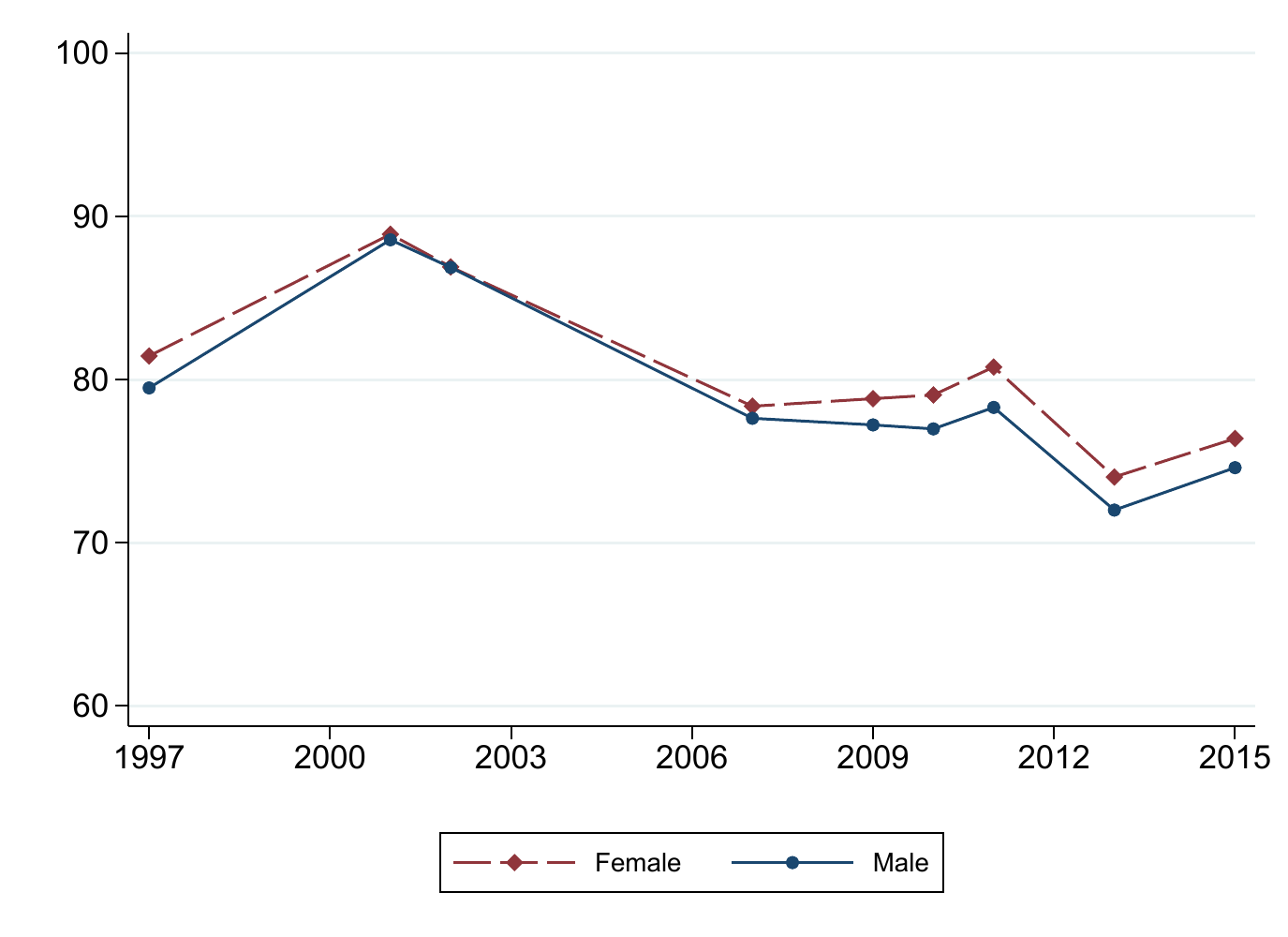}
	\end{subfigure}
	\hfill		
	\begin{subfigure}[t]{.48\textwidth}
		\caption*{Panel C. By educational attainment}\label{fig-fairness-educ}
		\centering
		\includegraphics[width=\linewidth]{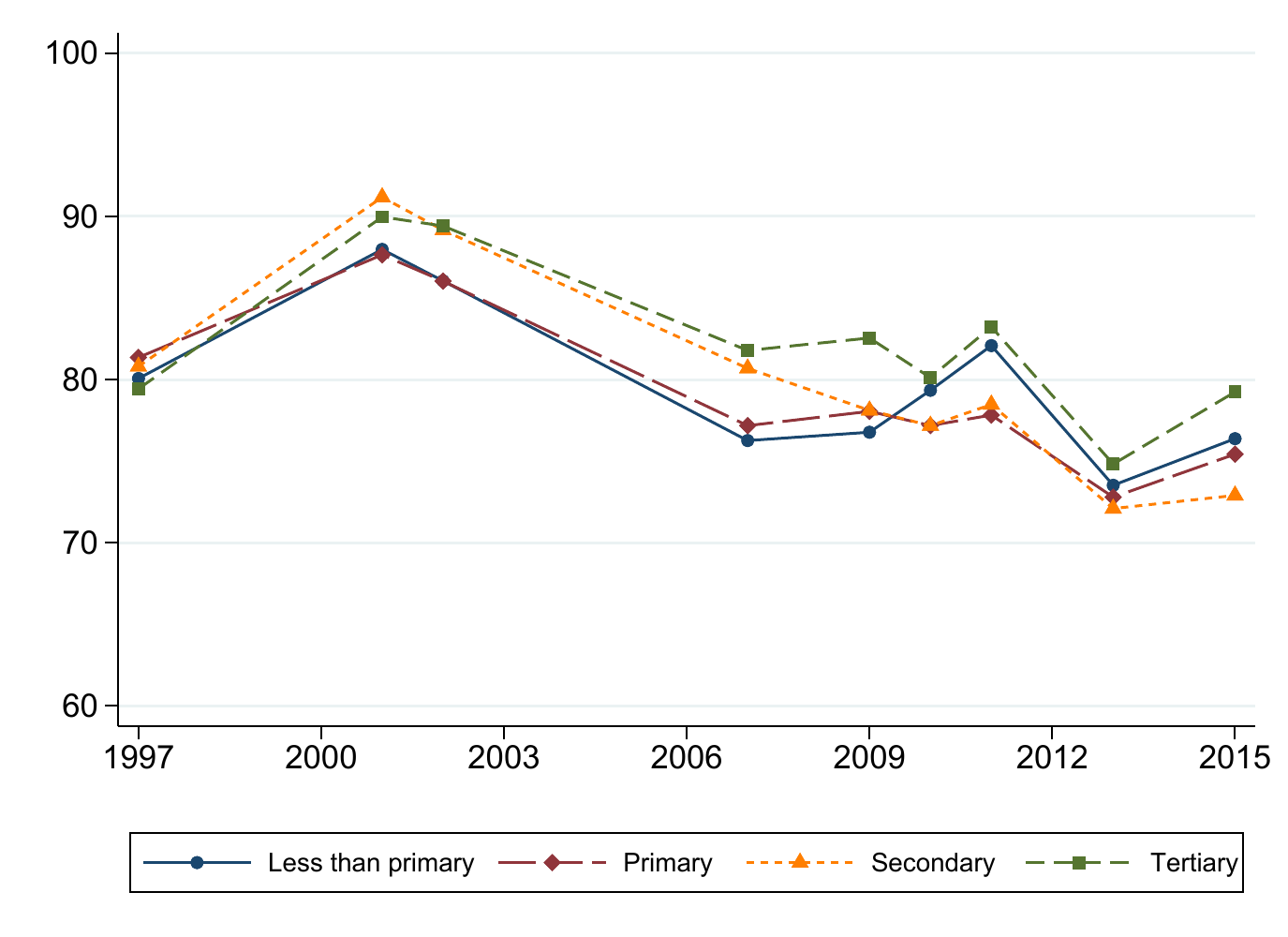}
	\end{subfigure}
	\hfill		
	\begin{subfigure}[t]{0.48\textwidth}
		\caption*{Panel D. By employment status}\label{fig-fairness-employ}
		\centering
		\includegraphics[width=\linewidth]{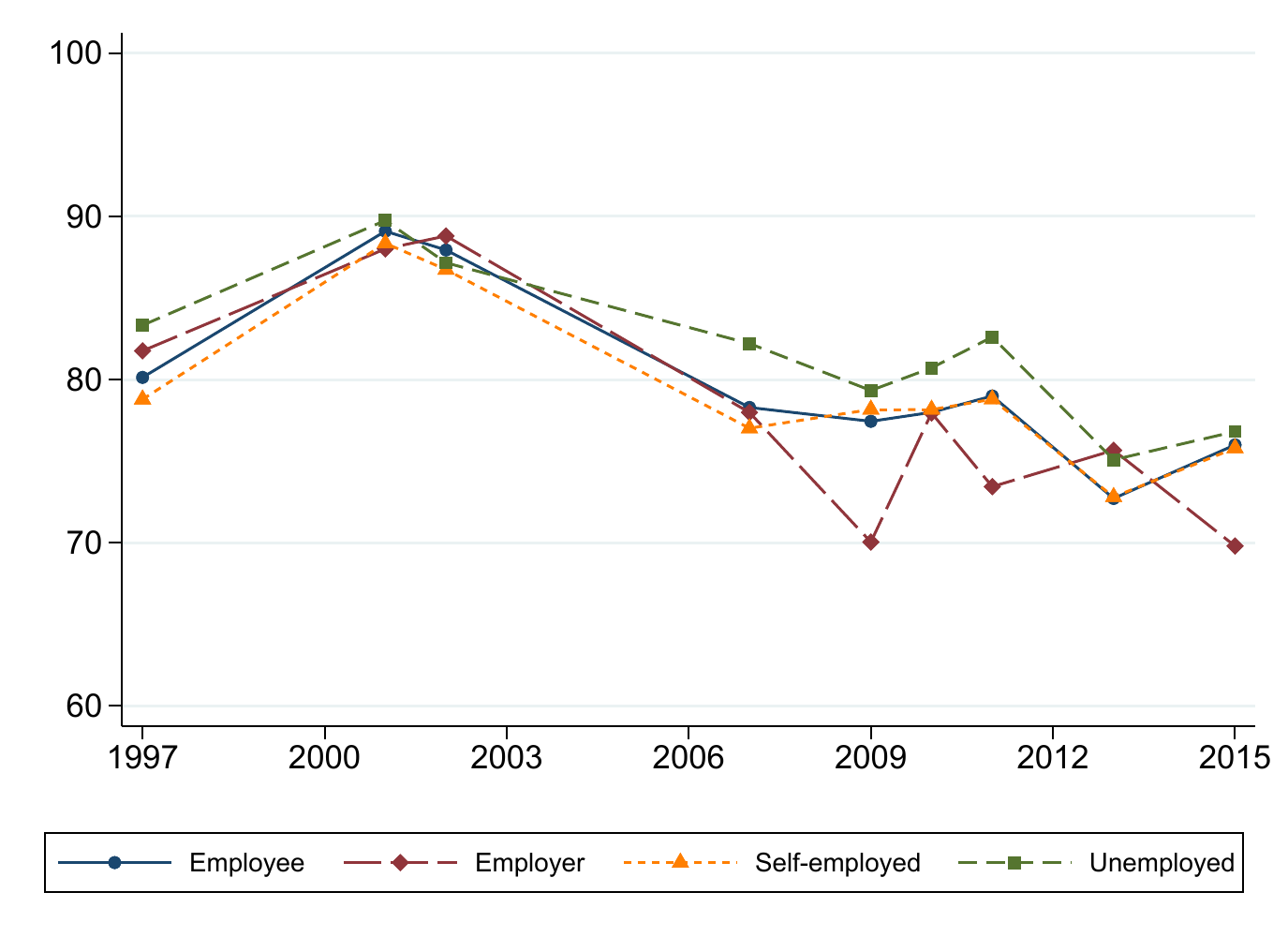}
	\end{subfigure}	
	\hfill				
	{\footnotesize
		\singlespacing \justify
		
		\textbf{Notes:} This figure shows the share of individuals who perceive the income distribution as unfair or very unfair according to their age, gender, maximum educational attainment, and employment status.

	}
\end{figure}

\clearpage 
\begin{figure}[htp]
	\caption{The evolution of fairness views and income inequality in Latin America}\label{fig-timeseries-gini-unfair}  \centering
	\centering
	\includegraphics[width=.75\linewidth]{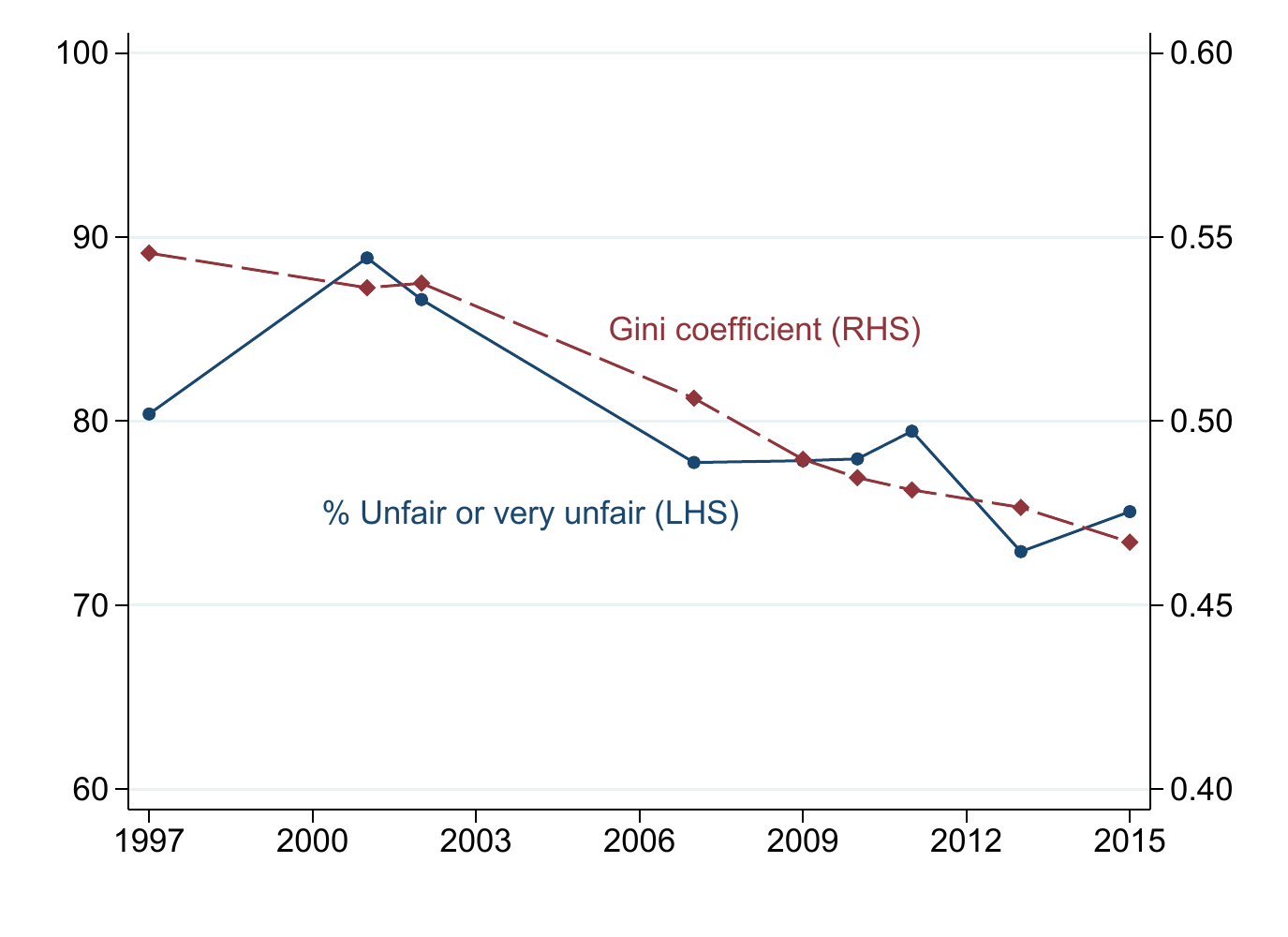}
	\hfill				
	{\footnotesize
		\singlespacing \justify
		
		\textbf{Notes:} This figure shows the evolution of the average Gini coefficient across countries in our sample (right-hand-side variable) and the fraction of the population who perceive the income distribution as unfair or very unfair (left-hand-side variable) over 1997--2015. To have a balanced panel of countries over time, we linearly extrapolated the Gini coefficient in years in which income microdata is not available (see Appendix \ref{sec_data}).
		
	}
\end{figure}

\clearpage 
\begin{table}[H]{\footnotesize
		\begin{center}
			\caption{Descriptive statistics of our sample} \label{tab-sum-stats}
			\begin{tabular}{lccc}
				\midrule
				&  Mean  &  Standard Dev.  & Observations \\
				& (1) & (2) & (3) \\
				\midrule
				\textbf{\hspace{-1em} Panel A. Sociodemographic} &   &   &  \\
				Age & 39.75 & 16.23 &         225,551  \\
				Male (\%) & 48.97 & 0.50 &         225,567  \\
				Married or civil union (\%) & 56.27 & 0.50 &         224,081  \\
				Catholic religion (\%) & 68.01 & 0.47 &         222,790  \\
				Ideology (10 = right-wing) & 5.48 & 2.64 &         131,980  \\
				\textbf{\hspace{-1em} Panel B. Education and Labor market} &   &   &  \\
				Literate (\%) & 90.31 & 0.30 &         224,056  \\
				Secondary education or more (\%) & 33.65 & 0.47 &         224,056  \\
				Parents with secondary education (\%) & 17.43 & 0.38 &         184,884  \\
				Economically active (\%) & 64.14 & 0.48 &         225,222  \\
				Unemployed (\% Labor Force) & 9.89 & 0.30 &         225,222  \\
				\textbf{\hspace{-1em} Panel C. Access to services} &   &   &  \\
				Access to a sewerage (\%) & 69.59 & 0.46 &         222,530  \\
				Access to running water (\%) & 88.83 & 0.31 &         204,340  \\
				\textbf{\hspace{-1em} Panel D. Asset ownership} &   &   &  \\
				Car (\%) & 28.21 & 0.45 &         222,338  \\
				Computer (\%) & 33.79 & 0.47 &         222,645  \\
				Fridge (\%) & 79.22 & 0.41 &         146,686  \\
				Homeowner (\%) & 73.92 & 0.44 &         223,603  \\
				Mobile (\%) & 80.61 & 0.40 &         172,253  \\
				Washing machine (\%) & 54.71 & 0.50 &         223,122  \\
				Landline (\%) & 42.28 & 0.49 &         222,968  \\
				\midrule
			\end{tabular}
		\end{center}
		\begin{singlespace} \vspace{-.5cm}
			\noindent \justify \textbf{Note:} This table shows summary statistics on our sample pooling data from all countries in our sample over 1997--2015.
		\end{singlespace}
	}
\end{table}

\clearpage 
\begin{table}[H]{\footnotesize
		\begin{center}
			\caption{Fairness views by population group} \label{tab-fairness-grp}
			\newcommand\w{1.70}
			\begin{tabular}{l@{}R{\w cm}R{\w cm}R{\w cm}R{\w cm}}
				\midrule					
				& \multicolumn{4}{c}{\% of individuals who believe income distribution is:} \\
				\cmidrule{2-5}  & Very unfair & Unfair & Fair & Very fair \\
				& (1) & (2) & (3) & (4) \\
				\midrule
				All & 28.2 & 51.6 & 17.3 & 2.9 \\
				\textbf{\hspace{-1em} Panel A. Gender} &   &   &   &  \\
				Female & 28.3 & 52.2 & 16.7 & 2.8 \\
				Male & 28.0 & 51.1 & 17.9 & 3.0 \\
				\textbf{\hspace{-1em} Panel B. Age group} &   &   &   &  \\
				15-24 & 25.2 & 52.0 & 19.7 & 3.1 \\
				25-40 & 28.5 & 51.3 & 17.2 & 3.0 \\
				41-64 & 29.5 & 51.7 & 16.0 & 2.8 \\
				65+ & 29.2 & 51.6 & 16.6 & 2.5 \\
				\textbf{\hspace{-1em} Panel C. Civil status} &   &   &   &  \\
				Married & 28.3 & 51.9 & 17.0 & 2.8 \\
				Not married & 27.9 & 51.4 & 17.7 & 3.1 \\
				\textbf{\hspace{-1em} Panel D. Religion} &   &   &   &  \\
				Catholic & 28.2 & 51.7 & 17.2 & 2.9 \\
				Not catholic & 28.0 & 51.5 & 17.5 & 3.0 \\
				\textbf{\hspace{-1em} Panel E. Education level} &   &   &   &  \\
				Less than Primary & 27.7 & 51.6 & 17.7 & 3.0 \\
				Complete Primary & 27.9 & 52.2 & 17.4 & 2.6 \\
				Complete Secondary & 29.1 & 53.2 & 15.0 & 2.7 \\
				Complete Tertiary & 29.0 & 50.8 & 17.1 & 3.1 \\
				\textbf{\hspace{-1em} Panel F. Type of employment} &   &   &   &  \\
				Employee & 28.3 & 51.5 & 17.2 & 2.9 \\
				Employer & 24.3 & 53.9 & 19.0 & 2.8 \\
				Self-employed & 28.0 & 51.4 & 17.5 & 3.1 \\
				Unemployed & 30.3 & 51.6 & 15.1 & 3.0 \\
				\textbf{\hspace{-1em} Panel E. Country} &   &   &   &  \\
				Argentina & 38.17 & 50.74 & 10.26 & 0.83 \\
				Bolivia & 18.01 & 56.13 & 23.39 & 2.48 \\
				Brazil & 31.95 & 53.71 & 12.85 & 1.49 \\
				Chile & 40.20 & 49.93 & 8.42 & 1.45 \\
				Colombia & 35.15 & 51.20 & 11.40 & 2.26 \\
				Costa Rica & 23.20 & 53.55 & 20.13 & 3.12 \\
				Dominican Rep. & 32.31 & 46.52 & 17.61 & 3.56 \\
				Ecuador & 21.45 & 47.46 & 27.58 & 3.51 \\
				El Salvador & 22.73 & 53.16 & 20.45 & 3.65 \\
				Guatemala & 28.29 & 51.34 & 16.70 & 3.66 \\
				Honduras & 28.87 & 53.42 & 14.33 & 3.38 \\
				Mexico & 32.15 & 49.75 & 15.32 & 2.78 \\
				Nicaragua & 18.69 & 51.88 & 24.33 & 5.11 \\
				Panama & 27.39 & 48.01 & 20.25 & 4.34 \\
				Paraguay & 38.31 & 48.80 & 10.95 & 1.93 \\
				Peru & 25.03 & 61.89 & 11.70 & 1.38 \\
				Uruguay & 18.22 & 57.51 & 22.64 & 1.64 \\
				Venezuela & 23.51 & 42.96 & 26.62 & 6.92 \\
				\midrule
			\end{tabular}
		\end{center}
		\begin{singlespace} \vspace{-.5cm}
			\noindent \justify \textbf{Note:} This table shows the fraction of individuals in our sample who perceive the income distribution as very unfair, unfair, fair, or very fair.
		\end{singlespace}
	}
\end{table}

\clearpage
\begin{table}[htpb!]{\footnotesize
		\begin{center}
			\caption{Logit regressions of unfairness perceptions (unfair) and individual characteristics} \label{unfair_logit}
			\newcommand\w{1.30}
			\begin{tabular}{l@{}lR{\w cm}@{}L{0.43cm}R{\w cm}@{}L{0.43cm}R{\w cm}@{}L{0.43cm}R{\w cm}@{}L{0.43cm}R{\w cm}@{}L{0.43cm}R{\w cm}@{}L{0.43cm}}
				\midrule
				&& 	\multicolumn{12}{c}{Dependent Variable: Believes income distribution is unfair or very unfair}  \\\cmidrule{3-14} 
				&& (1) && (2) && (3) && (4) && (5) && (6) \\	
				\midrule 
				Gini&&0.344&&0.347&&0.341&&0.338&&0.340&&0.355&\\
&&(0.220&)&(0.220&)&(0.221&)&(0.221&)&(0.220&)&(0.223&)\\
Age&&&&0.004&\sym{***}&0.004&\sym{***}&0.004&\sym{***}&0.004&\sym{***}&0.004&\sym{***}\\
&&&&(0.001&)&(0.000&)&(0.000&)&(0.000&)&(0.000&)\\
Age squared&&&&$-$0.000&\sym{***}&$-$0.000&\sym{***}&$-$0.000&\sym{***}&$-$0.000&\sym{***}&$-$0.000&\sym{***}\\
&&&&(0.000&)&(0.000&)&(0.000&)&(0.000&)&(0.000&)\\
Male&&&&$-$0.014&\sym{***}&$-$0.014&\sym{***}&$-$0.013&\sym{***}&$-$0.012&\sym{***}&$-$0.009&\sym{***}\\
&&&&(0.002&)&(0.002&)&(0.002&)&(0.002&)&(0.002&)\\
Married&&&&$-$0.000&&$-$0.000&&0.000&&0.000&&0.000&\\
&&&&(0.003&)&(0.003&)&(0.003&)&(0.003&)&(0.003&)\\
Finished HS&&&&&&0.004&&0.004&&0.006&&0.009&\sym{**}\\
&&&&&&(0.005&)&(0.005&)&(0.005&)&(0.005&)\\
Finished coll.&&&&&&0.010&\sym{*}&0.011&\sym{*}&0.016&\sym{***}&0.020&\sym{***}\\
&&&&&&(0.006&)&(0.006&)&(0.006&)&(0.006&)\\
In labor force&&&&&&&&$-$0.003&&$-$0.004&&$-$0.003&\\
&&&&&&&&(0.003&)&(0.003&)&(0.003&)\\
Unemployed&&&&&&&&0.021&\sym{***}&0.021&\sym{***}&0.021&\sym{***}\\
&&&&&&&&(0.005&)&(0.005&)&(0.005&)\\
Assets index&&&&&&&&&&$-$0.019&\sym{***}&$-$0.017&\sym{***}\\
&&&&&&&&&&(0.006&)&(0.006&)\\
Conservative&&&&&&&&&&&&0.001&\\
&&&&&&&&&&&&(0.002&)\\
Catholic&&&&&&&&&&&&$-$0.001&\\
&&&&&&&&&&&&(0.003&)\\
N&&167,436&&166,105&&164,662&&164,436&&164,436&&164,436&\\

				\midrule
			\end{tabular}
		\end{center}
		\begin{singlespace}  \vspace{-.5cm}
			\noindent \justify \textbf{Notes:} This table shows estimates of the relationship between an indicator that equals one for individuals who believe that the income distribution is unfair or very unfair and the Gini coefficient controlling for individuals' characteristics. Coefficients are estimated through Logit regressions and represent the marginal effects evaluated at the mean values of the rest of the variables. Observations are weighted by the individual's probability of being interviewed. All specifications include country and year fixed effects. $^{***}$, $^{**}$ and $^*$ denote significance at 10\%, 5\% and 1\% levels, respectively. Heteroskedasticity-robust standard errors clustered at the country-by-year level in parentheses. 
		\end{singlespace} 	
	}
\end{table}

\clearpage
\begin{table}[htpb!]{\footnotesize
		\begin{center}
			\caption{Logit regressions of unfairness perceptions (very unfair) and different inequality indicators} \label{unfair_ineq_logit}
			\newcommand\w{1.50}
			\begin{tabular}{l@{}lR{\w cm}@{}L{0.43cm}R{\w cm}@{}L{0.43cm}R{\w cm}@{}L{0.43cm}R{\w cm}@{}L{0.43cm}R{\w cm}@{}L{0.43cm}}
				\midrule
				&& 	\multicolumn{10}{c}{Dependent Variable: Believes income distribution is very unfair}  \\\cmidrule{3-12} 
				&& (1) && (2) && (3) && (4) && (5)  \\	
				\midrule 
				Gini (no zero income)&&0.708&\sym{***}&&&&&&&&\\
&&(0.223&)&&&&&&&&\\
Atkinson, A(1)&&&&0.341&\sym{**}&&&&&&\\
&&&&(0.164&)&&&&&&\\
Theil index&&&&&&0.246&\sym{***}&&&&\\
&&&&&&(0.080&)&&&&\\
Generalized entropy&&&&&&&&0.019&\sym{***}&&\\
&&&&&&&&(0.007&)&&\\
Absolute Gini&&&&&&&&&&$-$0.001&\\
&&&&&&&&&&(0.001&)\\
Age&&0.004&\sym{***}&0.004&\sym{***}&0.005&\sym{***}&0.005&\sym{***}&0.005&\sym{***}\\
&&(0.000&)&(0.000&)&(0.000&)&(0.000&)&(0.000&)\\
Age squared&&$-$0.000&\sym{***}&$-$0.000&\sym{***}&$-$0.000&\sym{***}&$-$0.000&\sym{***}&$-$0.000&\sym{***}\\
&&(0.000&)&(0.000&)&(0.000&)&(0.000&)&(0.000&)\\
Male&&0.001&&0.001&&0.001&&0.001&&0.001&\\
&&(0.003&)&(0.003&)&(0.003&)&(0.003&)&(0.003&)\\
Married&&$-$0.005&&$-$0.005&&$-$0.005&&$-$0.005&&$-$0.005&\sym{*}\\
&&(0.003&)&(0.003&)&(0.003&)&(0.003&)&(0.003&)\\
Finished HS&&$-$0.006&&$-$0.006&&$-$0.005&&$-$0.005&&$-$0.005&\\
&&(0.005&)&(0.005&)&(0.005&)&(0.005&)&(0.005&)\\
Finished coll.&&0.006&&0.006&&0.007&&0.007&&0.008&\\
&&(0.006&)&(0.006&)&(0.006&)&(0.006&)&(0.006&)\\
In labor force&&$-$0.002&&$-$0.002&&$-$0.002&&$-$0.002&&$-$0.002&\\
&&(0.003&)&(0.003&)&(0.003&)&(0.003&)&(0.003&)\\
Unemployed&&0.019&\sym{***}&0.020&\sym{***}&0.018&\sym{***}&0.018&\sym{***}&0.020&\sym{***}\\
&&(0.007&)&(0.007&)&(0.007&)&(0.007&)&(0.007&)\\
Assets index&&$-$0.007&&$-$0.007&&$-$0.007&&$-$0.006&&$-$0.007&\\
&&(0.006&)&(0.006&)&(0.006&)&(0.006&)&(0.006&)\\
Conservative&&$-$0.003&\sym{*}&$-$0.003&\sym{*}&$-$0.003&\sym{*}&$-$0.003&\sym{*}&$-$0.003&\sym{*}\\
&&(0.002&)&(0.002&)&(0.002&)&(0.002&)&(0.002&)\\
Catholic&&$-$0.011&\sym{***}&$-$0.011&\sym{***}&$-$0.011&\sym{***}&$-$0.011&\sym{***}&$-$0.011&\sym{***}\\
&&(0.004&)&(0.004&)&(0.004&)&(0.004&)&(0.004&)\\
N&&164,436&&164,436&&164,436&&164,436&&164,436&\\

				\midrule
			\end{tabular}
		\end{center}
		\begin{singlespace}  \vspace{-.5cm}
			\noindent \justify \textbf{Notes:} This table shows estimates of the relationship between an indicator that equals one for individuals who believe income distribution is unfair or very unfair and several inequality indicators controlling for individuals' characteristics. Coefficients are estimated through Logit regressions and represent the marginal effects evaluated at the mean values of the rest of the variables. Observations are weighted by the individual's probability of being interviewed. All specifications include country and year fixed effects. $^{***}$, $^{**}$ and $^*$ denote significance at 10\%, 5\% and 1\% levels, respectively. Heteroskedasticity-robust standard errors clustered at the country-by-year level in parentheses. 
			
		\end{singlespace} 	
	}
\end{table}

\begin{table}[htpb!]{\footnotesize
		\begin{center}
			\caption{OLS regressions of unfairness perceptions (very unfair) and individual characteristics} \label{very_unfair_lpm}
			\newcommand\w{1.30}
			\begin{tabular}{l@{}lR{\w cm}@{}L{0.43cm}R{\w cm}@{}L{0.43cm}R{\w cm}@{}L{0.43cm}R{\w cm}@{}L{0.43cm}R{\w cm}@{}L{0.43cm}R{\w cm}@{}L{0.43cm}}
				\midrule
				&& 	\multicolumn{12}{c}{Dependent Variable: Believes income distribution is very unfair}  \\\cmidrule{3-14} 
				&& (1) && (2) && (3) && (4) && (5) && (6) \\	
				\midrule 
				Gini&&0.621&\sym{***}&0.623&\sym{***}&0.622&\sym{***}&0.623&\sym{***}&0.623&\sym{***}&0.627&\sym{***}\\
&&(0.222&)&(0.222&)&(0.220&)&(0.220&)&(0.219&)&(0.218&)\\
Age&&&&0.004&\sym{***}&0.004&\sym{***}&0.004&\sym{***}&0.004&\sym{***}&0.004&\sym{***}\\
&&&&(0.000&)&(0.000&)&(0.000&)&(0.000&)&(0.000&)\\
Age squared&&&&$-$0.000&\sym{***}&$-$0.000&\sym{***}&$-$0.000&\sym{***}&$-$0.000&\sym{***}&$-$0.000&\sym{***}\\
&&&&(0.000&)&(0.000&)&(0.000&)&(0.000&)&(0.000&)\\
Male&&&&$-$0.002&&$-$0.002&&$-$0.001&&$-$0.001&&0.001&\\
&&&&(0.003&)&(0.003&)&(0.003&)&(0.003&)&(0.003&)\\
Married&&&&&&$-$0.006&\sym{**}&$-$0.005&\sym{*}&$-$0.005&\sym{*}&$-$0.005&\\
&&&&&&(0.003&)&(0.003&)&(0.003&)&(0.003&)\\
Finished HS&&&&&&$-$0.010&\sym{**}&$-$0.010&\sym{**}&$-$0.009&\sym{*}&$-$0.007&\\
&&&&&&(0.005&)&(0.005&)&(0.005&)&(0.005&)\\
Finished coll.&&&&&&$-$0.002&&$-$0.001&&0.001&&0.005&\\
&&&&&&(0.007&)&(0.007&)&(0.006&)&(0.006&)\\
In labor force&&&&&&&&$-$0.003&&$-$0.003&&$-$0.002&\\
&&&&&&&&(0.003&)&(0.003&)&(0.003&)\\
Unemployed&&&&&&&&0.020&\sym{***}&0.020&\sym{***}&0.019&\sym{***}\\
&&&&&&&&(0.007&)&(0.007&)&(0.007&)\\
Assets index&&&&&&&&&&$-$0.009&&$-$0.007&\\
&&&&&&&&&&(0.006&)&(0.006&)\\
Conservative&&&&&&&&&&&&$-$0.003&\sym{*}\\
&&&&&&&&&&&&(0.002&)\\
Catholic&&&&&&&&&&&&$-$0.011&\sym{***}\\
&&&&&&&&&&&&(0.004&)\\
N&&167,436&&167,420&&164,662&&164,436&&164,436&&164,436&\\

				\midrule
			\end{tabular}
		\end{center}
		\begin{singlespace}  \vspace{-.5cm}
			\noindent \justify \textbf{Notes:} This table presents estimates of the correlation between a dummy variable that indicates if the individual believes income distribution is unfair or very unfair and the Gini coefficient controlling for individuals' characteristics. Coefficients are estimated through a linear probability model. Observations are weighted by the individual's probability of being interviewed. All specifications include country and year fixed effects. $^{***}$, $^{**}$ and $^*$ denote significance at 10\%, 5\% and 1\% levels, respectively. Heteroskedasticity-robust standard errors clustered at the country-by-year level in parentheses. 
		\end{singlespace} 	
	}
\end{table}
 	
	\clearpage
\section{Data Appendix} \label{sec_data} 

\setcounter{table}{0}
\setcounter{figure}{0}
\renewcommand{\thetable}{B\arabic{table}}
\renewcommand{\thefigure}{B\arabic{figure}}

The figures presented in this paper are based on two harmonization projects, known as Latinobar\'ometro and SEDLAC (Socio-Economic Database for Latin America and the Caribbean). In this Appendix, we describe how we make both sources compatible. 

Our perceptions data come from Latinobar\'ometro, which has conducted opinion surveys in 18 LA countries since the 1990s, interviewing about 1,200 individuals per country about individuals' socioeconomic background, and preferences towards political and social issues. Unfortunately, not all years contain questions about individuals' fairness perceptions. The survey was designed to be representative of the voting-age population at the national level (in most LA countries, individuals aged over 18). In Table \ref{tab-coverage} we show what percentage of the voting-age population is represented by the survey in each country for all the years in which the fairness question is available.

\begin{table}[htbp]{\footnotesize
		\begin{center}
			\caption{Coverage of each country's population in Latinobar\'ometro overtime (in \%)}\label{tab-coverage}
			\begin{tabular}{lrrrrrrrrr}
				\midrule
				& 1997 & 2001 & 2002 & 2007 & 2009 & 2010 & 2011 & 2013 & 2015 \\
				\midrule
				Argentina &         68  &         75  &         75  &       100  &       100  &       100  &       100  &       100  &       100  \\
				Bolivia &         32  &         52  &       100  &       100  &       100  &       100  &       100  &       100  &       100  \\
				Brazil &         12  &       100  &       100  &       100  &       100  &       100  &       100  &       100  &       100  \\
				Chile &         70  &         70  &         70  &       100  &       100  &       100  &       100  &       100  &       100  \\
				Colombia &         25  &         71  &         51  &       100  &       100  &       100  &       100  &       100  &       100  \\
				Costa Rica &       100  &       100  &       100  &       100  &       100  &       100  &       100  &       100  &       100  \\
				Dominican Republic & \multicolumn{1}{c}{ N/A } & \multicolumn{1}{c}{ N/A } & \multicolumn{1}{c}{ N/A } &       100  &       100  &       100  &       100  &       100  &       100  \\
				Ecuador &         97  &         97  &       100  &       100  &       100  &       100  &       100  &       100  &       100  \\
				El Salvador &         65  &       100  &       100  &       100  &       100  &       100  &       100  &       100  &       100  \\
				Guatemala &       100  &       100  &       100  &         97  &       100  &       100  &       100  &       100  &       100  \\
				Honduras &       100  &       100  &       100  &         98  &       100  &         99  &         99  &         99  &         99  \\
				Mexico &         93  &         88  &         95  &       100  &       100  &       100  &       100  &       100  &       100  \\
				Nicaragua &       100  &       100  &       100  &       100  &       100  &       100  &       100  &       100  &       100  \\
				Panama &       100  &       100  &       100  &         99  &         99  &         99  &         99  &         99  &         99  \\
				Paraguay &         46  &         46  &         46  &       100  &       100  &       100  &       100  &       100  &       100  \\
				Peru &         52  &         52  &       100  &       100  &       100  &       100  &       100  &       100  &       100  \\
				Uruguay &         80  &         80  &         80  &       100  &       100  &       100  &       100  &       100  &       100  \\
				Venezuela &       100  &       100  &       100  &       100  &         93  &       100  &       100  &       100  &       100  \\
				Weighted average &         68  &         86  &         91  &       100  &       100  &       100  &       100  &       100  &       100  \\
				\midrule
			\end{tabular}
		\end{center}
	}
\end{table}

Since our goal is to analyze how unfairness perceptions evolved vis-\`a-vis changes in income inequality, we put a lot of effort into getting income inequality data for each data point for which we have perceptions data available. We made two partial fixes to increase the number of observations available (without pushing the data too much). First, we filled the data gaps using household surveys of relatively close years in which previously unused data were available (see Appendix Table \ref{tab-circa}). For instance, Chile conducts household surveys on average every two years. In 1997, there is perceptions data available, but no data on income inequality. Therefore, we use the inequality data from an adjacent year (1998). As noted previously, we only use data from close years if the data from the adjacent year correspond to a year in which the perceptions question was not asked (and therefore, inequality data are not needed in that year).

\begin{table}[htbp]{\footnotesize
		\begin{center}
			\caption{Circa years used to fill data gaps}\label{tab-circa}
			\begin{tabular}{lcc}
				\midrule
				Country & Year without household data & Data point used instead \\
				\midrule
				Chile & 1997 & 1998 \\
				Chile & 2001 & 2000 \\
				Chile & 2002 & 2003 \\
				Chile & 2007 & 2006 \\
				Colombia & 2007 & 2008 \\
				Ecuador & 2002 & 2003 \\
				El Salvador & 1997 & 1998 \\
				Guatemala & 2001 & 2000 \\
				Guatemala & 2015 & 2014 \\
				Mexico & 1997 & 1998 \\
				Mexico & 2001 & 2000 \\
				Mexico & 2007 & 2006 \\
				Mexico & 2009 & 2008 \\
				Mexico & 2011 & 2012 \\
				Mexico & 2015 & 2014 \\
				Nicaragua & 1997 & 1998 \\
				Nicaragua & 2007 & 2005 \\
				Nicaragua & 2015 & 2014 \\
				Venezuela & 2013 & 2012 \\
				\midrule
			\end{tabular}
		\end{center}
		\begin{singlespace}  \vspace{-.5cm}
			\noindent \justify 
		\end{singlespace} 	
	}
\end{table}

Our second partial fix involves interpolating inequality indicators for some years. For some countries, a few years had perceptions data available but no comparable household survey over time and no close year available. In this case, and to analyze the same set of countries every year, interpolation was applied to the inequality indicators (see Appendix Table \ref{tab-interp}).

\begin{table}[htbp]{\footnotesize
		\begin{center}
			\caption{Years in which inequality indicators were calculated with a linear interpolation}\label{tab-interp}
			\begin{tabular}{ll}
				\midrule
				Country & Years interpolated \\
				\midrule
				Argentina & 1997, 2001, and 2002 \\
				Bolivia & 2010 \\
				Brazil & 2010 \\
				Chile & 2010 \\
				Colombia & 1997 \\
				Costa Rica & 1997, 2001, 2002, 2007, and 2009 \\
				Ecuador & 1997, 2001 \\
				Guatemala & 1997, 2002, 2009, 2010, and 2013 \\
				Mexico & 2013 \\
				Nicaragua & 2002, 2010, 2011, and 2013 \\
				Panama & 1997, 2001, 2002, and 2007 \\
				Peru & 1997, 2001, 2002 \\
				Venezuela & 2015 \\
				\midrule
			\end{tabular}
		\end{center}
		\begin{singlespace}  \vspace{-.5cm}
			\noindent \justify 
		\end{singlespace} 	
	}
\end{table}

Overall, the years in which income inequality was calculated using linear interpolations represent a relatively small share of the total data points (17\% of total). The majority of our inequality data points (69\%) are calculated using a household survey from the same year in which the perceptions polls were conducted, while the remaining 14\% of our inequality indicators are calculated using household surveys from adjacent years. Table \ref{tab-summ-data} summarizes the data sources used in years perceptions data are available.

\begin{table}[htbp]{\footnotesize
		\begin{center}
			\caption{Summary of the data used in every country-year}\label{tab-summ-data}
			\begin{tabular}{r|r|llllllll}
				\multicolumn{1}{r}{} & \multicolumn{1}{r}{1997} & \multicolumn{1}{r}{2001} & \multicolumn{1}{r}{2002} & \multicolumn{1}{r}{2007} & \multicolumn{1}{r}{2009} & \multicolumn{1}{r}{2010} & \multicolumn{1}{r}{2011} & \multicolumn{1}{r}{2013} & \multicolumn{1}{r}{2015} \\
				\cmidrule{2-10}    \multicolumn{1}{l|}{Argentina} & \cellcolor[rgb]{ .608,  .733,  .349}\textcolor[rgb]{ .608,  .733,  .349}{3} & \multicolumn{1}{r|}{\cellcolor[rgb]{ .608,  .733,  .349}\textcolor[rgb]{ .608,  .733,  .349}{3}} & \multicolumn{1}{r|}{\cellcolor[rgb]{ .608,  .733,  .349}\textcolor[rgb]{ .608,  .733,  .349}{3}} & \multicolumn{1}{r|}{\cellcolor[rgb]{ .31,  .506,  .741}\textcolor[rgb]{ .31,  .506,  .741}{1}} & \multicolumn{1}{r|}{\cellcolor[rgb]{ .31,  .506,  .741}\textcolor[rgb]{ .31,  .506,  .741}{1}} & \multicolumn{1}{r|}{\cellcolor[rgb]{ .31,  .506,  .741}\textcolor[rgb]{ .31,  .506,  .741}{1}} & \multicolumn{1}{r|}{\cellcolor[rgb]{ .31,  .506,  .741}\textcolor[rgb]{ .31,  .506,  .741}{1}} & \multicolumn{1}{r|}{\cellcolor[rgb]{ .31,  .506,  .741}\textcolor[rgb]{ .31,  .506,  .741}{1}} & \multicolumn{1}{r|}{\cellcolor[rgb]{ .969,  .588,  .275}\textcolor[rgb]{ .969,  .588,  .275}{2}} \\
				\cmidrule{2-10}    \multicolumn{1}{l|}{Bolivia} & \cellcolor[rgb]{ .31,  .506,  .741}\textcolor[rgb]{ .31,  .506,  .741}{1} & \multicolumn{1}{r|}{\cellcolor[rgb]{ .31,  .506,  .741}\textcolor[rgb]{ .31,  .506,  .741}{1}} & \multicolumn{1}{r|}{\cellcolor[rgb]{ .31,  .506,  .741}\textcolor[rgb]{ .31,  .506,  .741}{1}} & \multicolumn{1}{r|}{\cellcolor[rgb]{ .31,  .506,  .741}\textcolor[rgb]{ .31,  .506,  .741}{1}} & \multicolumn{1}{r|}{\cellcolor[rgb]{ .31,  .506,  .741}\textcolor[rgb]{ .31,  .506,  .741}{1}} & \multicolumn{1}{r|}{\cellcolor[rgb]{ .608,  .733,  .349}\textcolor[rgb]{ .608,  .733,  .349}{3}} & \multicolumn{1}{r|}{\cellcolor[rgb]{ .31,  .506,  .741}\textcolor[rgb]{ .31,  .506,  .741}{1}} & \multicolumn{1}{r|}{\cellcolor[rgb]{ .31,  .506,  .741}\textcolor[rgb]{ .31,  .506,  .741}{1}} & \multicolumn{1}{r|}{\cellcolor[rgb]{ .31,  .506,  .741}\textcolor[rgb]{ .31,  .506,  .741}{1}} \\
				\cmidrule{2-10}    \multicolumn{1}{l|}{Brazil} & \cellcolor[rgb]{ .31,  .506,  .741}\textcolor[rgb]{ .31,  .506,  .741}{1} & \multicolumn{1}{r|}{\cellcolor[rgb]{ .31,  .506,  .741}\textcolor[rgb]{ .31,  .506,  .741}{1}} & \multicolumn{1}{r|}{\cellcolor[rgb]{ .31,  .506,  .741}\textcolor[rgb]{ .31,  .506,  .741}{1}} & \multicolumn{1}{r|}{\cellcolor[rgb]{ .31,  .506,  .741}\textcolor[rgb]{ .31,  .506,  .741}{1}} & \multicolumn{1}{r|}{\cellcolor[rgb]{ .31,  .506,  .741}\textcolor[rgb]{ .31,  .506,  .741}{1}} & \multicolumn{1}{r|}{\cellcolor[rgb]{ .608,  .733,  .349}\textcolor[rgb]{ .608,  .733,  .349}{3}} & \multicolumn{1}{r|}{\cellcolor[rgb]{ .31,  .506,  .741}\textcolor[rgb]{ .31,  .506,  .741}{1}} & \multicolumn{1}{r|}{\cellcolor[rgb]{ .31,  .506,  .741}\textcolor[rgb]{ .31,  .506,  .741}{1}} & \multicolumn{1}{r|}{\cellcolor[rgb]{ .31,  .506,  .741}\textcolor[rgb]{ .31,  .506,  .741}{1}} \\
				\cmidrule{2-10}    \multicolumn{1}{l|}{Chile} & \cellcolor[rgb]{ .969,  .588,  .275}\textcolor[rgb]{ .969,  .588,  .275}{2} & \multicolumn{1}{r|}{\cellcolor[rgb]{ .969,  .588,  .275}\textcolor[rgb]{ .969,  .588,  .275}{2}} & \multicolumn{1}{r|}{\cellcolor[rgb]{ .969,  .588,  .275}\textcolor[rgb]{ .969,  .588,  .275}{2}} & \multicolumn{1}{r|}{\cellcolor[rgb]{ .969,  .588,  .275}\textcolor[rgb]{ .969,  .588,  .275}{2}} & \multicolumn{1}{r|}{\cellcolor[rgb]{ .31,  .506,  .741}\textcolor[rgb]{ .31,  .506,  .741}{1}} & \multicolumn{1}{r|}{\cellcolor[rgb]{ .608,  .733,  .349}\textcolor[rgb]{ .608,  .733,  .349}{3}} & \multicolumn{1}{r|}{\cellcolor[rgb]{ .31,  .506,  .741}\textcolor[rgb]{ .31,  .506,  .741}{1}} & \multicolumn{1}{r|}{\cellcolor[rgb]{ .31,  .506,  .741}\textcolor[rgb]{ .31,  .506,  .741}{1}} & \multicolumn{1}{r|}{\cellcolor[rgb]{ .31,  .506,  .741}\textcolor[rgb]{ .31,  .506,  .741}{1}} \\
				\cmidrule{2-10}    \multicolumn{1}{l|}{Colombia} & \cellcolor[rgb]{ .608,  .733,  .349}\textcolor[rgb]{ .608,  .733,  .349}{3} & \multicolumn{1}{r|}{\cellcolor[rgb]{ .31,  .506,  .741}\textcolor[rgb]{ .31,  .506,  .741}{1}} & \multicolumn{1}{r|}{\cellcolor[rgb]{ .31,  .506,  .741}\textcolor[rgb]{ .31,  .506,  .741}{1}} & \multicolumn{1}{r|}{\cellcolor[rgb]{ .969,  .588,  .275}\textcolor[rgb]{ .969,  .588,  .275}{2}} & \multicolumn{1}{r|}{\cellcolor[rgb]{ .31,  .506,  .741}\textcolor[rgb]{ .31,  .506,  .741}{1}} & \multicolumn{1}{r|}{\cellcolor[rgb]{ .31,  .506,  .741}\textcolor[rgb]{ .31,  .506,  .741}{1}} & \multicolumn{1}{r|}{\cellcolor[rgb]{ .31,  .506,  .741}\textcolor[rgb]{ .31,  .506,  .741}{1}} & \multicolumn{1}{r|}{\cellcolor[rgb]{ .31,  .506,  .741}\textcolor[rgb]{ .31,  .506,  .741}{1}} & \multicolumn{1}{r|}{\cellcolor[rgb]{ .31,  .506,  .741}\textcolor[rgb]{ .31,  .506,  .741}{1}} \\
				\cmidrule{2-10}    \multicolumn{1}{l|}{Costa Rica} & \cellcolor[rgb]{ .608,  .733,  .349}\textcolor[rgb]{ .608,  .733,  .349}{3} & \multicolumn{1}{r|}{\cellcolor[rgb]{ .608,  .733,  .349}\textcolor[rgb]{ .608,  .733,  .349}{3}} & \multicolumn{1}{r|}{\cellcolor[rgb]{ .608,  .733,  .349}\textcolor[rgb]{ .608,  .733,  .349}{3}} & \multicolumn{1}{r|}{\cellcolor[rgb]{ .608,  .733,  .349}\textcolor[rgb]{ .608,  .733,  .349}{3}} & \multicolumn{1}{r|}{\cellcolor[rgb]{ .608,  .733,  .349}\textcolor[rgb]{ .608,  .733,  .349}{3}} & \multicolumn{1}{r|}{\cellcolor[rgb]{ .31,  .506,  .741}\textcolor[rgb]{ .31,  .506,  .741}{1}} & \multicolumn{1}{r|}{\cellcolor[rgb]{ .31,  .506,  .741}\textcolor[rgb]{ .31,  .506,  .741}{1}} & \multicolumn{1}{r|}{\cellcolor[rgb]{ .31,  .506,  .741}\textcolor[rgb]{ .31,  .506,  .741}{1}} & \multicolumn{1}{r|}{\cellcolor[rgb]{ .31,  .506,  .741}\textcolor[rgb]{ .31,  .506,  .741}{1}} \\
				\cmidrule{2-10}    \multicolumn{1}{l|}{Dominican Rep.} & \cellcolor[rgb]{ .749,  .749,  .749}\textcolor[rgb]{ .749,  .749,  .749}{0} & \multicolumn{1}{r|}{\cellcolor[rgb]{ .31,  .506,  .741}\textcolor[rgb]{ .31,  .506,  .741}{1}} & \multicolumn{1}{r|}{\cellcolor[rgb]{ .31,  .506,  .741}\textcolor[rgb]{ .31,  .506,  .741}{1}} & \multicolumn{1}{r|}{\cellcolor[rgb]{ .31,  .506,  .741}\textcolor[rgb]{ .31,  .506,  .741}{1}} & \multicolumn{1}{r|}{\cellcolor[rgb]{ .31,  .506,  .741}\textcolor[rgb]{ .31,  .506,  .741}{1}} & \multicolumn{1}{r|}{\cellcolor[rgb]{ .31,  .506,  .741}\textcolor[rgb]{ .31,  .506,  .741}{1}} & \multicolumn{1}{r|}{\cellcolor[rgb]{ .31,  .506,  .741}\textcolor[rgb]{ .31,  .506,  .741}{1}} & \multicolumn{1}{r|}{\cellcolor[rgb]{ .31,  .506,  .741}\textcolor[rgb]{ .31,  .506,  .741}{1}} & \multicolumn{1}{r|}{\cellcolor[rgb]{ .31,  .506,  .741}\textcolor[rgb]{ .31,  .506,  .741}{1}} \\
				\cmidrule{2-10}    \multicolumn{1}{l|}{Ecuador} & \cellcolor[rgb]{ .608,  .733,  .349}\textcolor[rgb]{ .608,  .733,  .349}{3} & \multicolumn{1}{r|}{\cellcolor[rgb]{ .608,  .733,  .349}\textcolor[rgb]{ .608,  .733,  .349}{3}} & \multicolumn{1}{r|}{\cellcolor[rgb]{ .969,  .588,  .275}\textcolor[rgb]{ .969,  .588,  .275}{2}} & \multicolumn{1}{r|}{\cellcolor[rgb]{ .31,  .506,  .741}\textcolor[rgb]{ .31,  .506,  .741}{1}} & \multicolumn{1}{r|}{\cellcolor[rgb]{ .31,  .506,  .741}\textcolor[rgb]{ .31,  .506,  .741}{1}} & \multicolumn{1}{r|}{\cellcolor[rgb]{ .31,  .506,  .741}\textcolor[rgb]{ .31,  .506,  .741}{1}} & \multicolumn{1}{r|}{\cellcolor[rgb]{ .31,  .506,  .741}\textcolor[rgb]{ .31,  .506,  .741}{1}} & \multicolumn{1}{r|}{\cellcolor[rgb]{ .31,  .506,  .741}\textcolor[rgb]{ .31,  .506,  .741}{1}} & \multicolumn{1}{r|}{\cellcolor[rgb]{ .31,  .506,  .741}\textcolor[rgb]{ .31,  .506,  .741}{1}} \\
				\cmidrule{2-10}    \multicolumn{1}{l|}{El Salvador} & \cellcolor[rgb]{ .31,  .506,  .741}\textcolor[rgb]{ .31,  .506,  .741}{1} & \multicolumn{1}{r|}{\cellcolor[rgb]{ .31,  .506,  .741}\textcolor[rgb]{ .31,  .506,  .741}{1}} & \multicolumn{1}{r|}{\cellcolor[rgb]{ .31,  .506,  .741}\textcolor[rgb]{ .31,  .506,  .741}{1}} & \multicolumn{1}{r|}{\cellcolor[rgb]{ .31,  .506,  .741}\textcolor[rgb]{ .31,  .506,  .741}{1}} & \multicolumn{1}{r|}{\cellcolor[rgb]{ .31,  .506,  .741}\textcolor[rgb]{ .31,  .506,  .741}{1}} & \multicolumn{1}{r|}{\cellcolor[rgb]{ .31,  .506,  .741}\textcolor[rgb]{ .31,  .506,  .741}{1}} & \multicolumn{1}{r|}{\cellcolor[rgb]{ .31,  .506,  .741}\textcolor[rgb]{ .31,  .506,  .741}{1}} & \multicolumn{1}{r|}{\cellcolor[rgb]{ .31,  .506,  .741}\textcolor[rgb]{ .31,  .506,  .741}{1}} & \multicolumn{1}{r|}{\cellcolor[rgb]{ .31,  .506,  .741}\textcolor[rgb]{ .31,  .506,  .741}{1}} \\
				\cmidrule{2-10}    \multicolumn{1}{l|}{Guatemala} & \cellcolor[rgb]{ .969,  .588,  .275}\textcolor[rgb]{ .969,  .588,  .275}{2} & \multicolumn{1}{r|}{\cellcolor[rgb]{ .969,  .588,  .275}\textcolor[rgb]{ .969,  .588,  .275}{2}} & \multicolumn{1}{r|}{\cellcolor[rgb]{ .608,  .733,  .349}\textcolor[rgb]{ .608,  .733,  .349}{3}} & \multicolumn{1}{r|}{\cellcolor[rgb]{ .969,  .588,  .275}\textcolor[rgb]{ .969,  .588,  .275}{2}} & \multicolumn{1}{r|}{\cellcolor[rgb]{ .608,  .733,  .349}\textcolor[rgb]{ .608,  .733,  .349}{3}} & \multicolumn{1}{r|}{\cellcolor[rgb]{ .608,  .733,  .349}\textcolor[rgb]{ .608,  .733,  .349}{3}} & \multicolumn{1}{r|}{\cellcolor[rgb]{ .31,  .506,  .741}\textcolor[rgb]{ .31,  .506,  .741}{1}} & \multicolumn{1}{r|}{\cellcolor[rgb]{ .608,  .733,  .349}\textcolor[rgb]{ .608,  .733,  .349}{3}} & \multicolumn{1}{r|}{\cellcolor[rgb]{ .969,  .588,  .275}\textcolor[rgb]{ .969,  .588,  .275}{2}} \\
				\cmidrule{2-10}    \multicolumn{1}{l|}{Honduras} & \cellcolor[rgb]{ .31,  .506,  .741}\textcolor[rgb]{ .31,  .506,  .741}{1} & \multicolumn{1}{r|}{\cellcolor[rgb]{ .31,  .506,  .741}\textcolor[rgb]{ .31,  .506,  .741}{1}} & \multicolumn{1}{r|}{\cellcolor[rgb]{ .31,  .506,  .741}\textcolor[rgb]{ .31,  .506,  .741}{1}} & \multicolumn{1}{r|}{\cellcolor[rgb]{ .31,  .506,  .741}\textcolor[rgb]{ .31,  .506,  .741}{1}} & \multicolumn{1}{r|}{\cellcolor[rgb]{ .31,  .506,  .741}\textcolor[rgb]{ .31,  .506,  .741}{1}} & \multicolumn{1}{r|}{\cellcolor[rgb]{ .31,  .506,  .741}\textcolor[rgb]{ .31,  .506,  .741}{1}} & \multicolumn{1}{r|}{\cellcolor[rgb]{ .31,  .506,  .741}\textcolor[rgb]{ .31,  .506,  .741}{1}} & \multicolumn{1}{r|}{\cellcolor[rgb]{ .31,  .506,  .741}\textcolor[rgb]{ .31,  .506,  .741}{1}} & \multicolumn{1}{r|}{\cellcolor[rgb]{ .31,  .506,  .741}\textcolor[rgb]{ .31,  .506,  .741}{1}} \\
				\cmidrule{2-10}    \multicolumn{1}{l|}{Mexico} & \cellcolor[rgb]{ .969,  .588,  .275}\textcolor[rgb]{ .969,  .588,  .275}{2} & \multicolumn{1}{r|}{\cellcolor[rgb]{ .969,  .588,  .275}\textcolor[rgb]{ .969,  .588,  .275}{2}} & \multicolumn{1}{r|}{\cellcolor[rgb]{ .31,  .506,  .741}\textcolor[rgb]{ .31,  .506,  .741}{1}} & \multicolumn{1}{r|}{\cellcolor[rgb]{ .969,  .588,  .275}\textcolor[rgb]{ .969,  .588,  .275}{2}} & \multicolumn{1}{r|}{\cellcolor[rgb]{ .969,  .588,  .275}\textcolor[rgb]{ .969,  .588,  .275}{2}} & \multicolumn{1}{r|}{\cellcolor[rgb]{ .31,  .506,  .741}\textcolor[rgb]{ .31,  .506,  .741}{1}} & \multicolumn{1}{r|}{\cellcolor[rgb]{ .969,  .588,  .275}\textcolor[rgb]{ .969,  .588,  .275}{2}} & \multicolumn{1}{r|}{\cellcolor[rgb]{ .608,  .733,  .349}\textcolor[rgb]{ .608,  .733,  .349}{3}} & \multicolumn{1}{r|}{\cellcolor[rgb]{ .969,  .588,  .275}\textcolor[rgb]{ .969,  .588,  .275}{2}} \\
				\cmidrule{2-10}    \multicolumn{1}{l|}{Nicaragua} & \cellcolor[rgb]{ .969,  .588,  .275}\textcolor[rgb]{ .969,  .588,  .275}{2} & \multicolumn{1}{r|}{\cellcolor[rgb]{ .31,  .506,  .741}\textcolor[rgb]{ .31,  .506,  .741}{1}} & \multicolumn{1}{r|}{\cellcolor[rgb]{ .608,  .733,  .349}\textcolor[rgb]{ .608,  .733,  .349}{3}} & \multicolumn{1}{r|}{\cellcolor[rgb]{ .969,  .588,  .275}\textcolor[rgb]{ .969,  .588,  .275}{2}} & \multicolumn{1}{r|}{\cellcolor[rgb]{ .31,  .506,  .741}\textcolor[rgb]{ .31,  .506,  .741}{1}} & \multicolumn{1}{r|}{\cellcolor[rgb]{ .608,  .733,  .349}\textcolor[rgb]{ .608,  .733,  .349}{3}} & \multicolumn{1}{r|}{\cellcolor[rgb]{ .608,  .733,  .349}\textcolor[rgb]{ .608,  .733,  .349}{3}} & \multicolumn{1}{r|}{\cellcolor[rgb]{ .608,  .733,  .349}\textcolor[rgb]{ .608,  .733,  .349}{3}} & \multicolumn{1}{r|}{\cellcolor[rgb]{ .969,  .588,  .275}\textcolor[rgb]{ .969,  .588,  .275}{2}} \\
				\cmidrule{2-10}    \multicolumn{1}{l|}{Panama} & \cellcolor[rgb]{ .608,  .733,  .349}\textcolor[rgb]{ .608,  .733,  .349}{3} & \multicolumn{1}{r|}{\cellcolor[rgb]{ .608,  .733,  .349}\textcolor[rgb]{ .608,  .733,  .349}{3}} & \multicolumn{1}{r|}{\cellcolor[rgb]{ .608,  .733,  .349}\textcolor[rgb]{ .608,  .733,  .349}{3}} & \multicolumn{1}{r|}{\cellcolor[rgb]{ .608,  .733,  .349}\textcolor[rgb]{ .608,  .733,  .349}{3}} & \multicolumn{1}{r|}{\cellcolor[rgb]{ .31,  .506,  .741}\textcolor[rgb]{ .31,  .506,  .741}{1}} & \multicolumn{1}{r|}{\cellcolor[rgb]{ .31,  .506,  .741}\textcolor[rgb]{ .31,  .506,  .741}{1}} & \multicolumn{1}{r|}{\cellcolor[rgb]{ .31,  .506,  .741}\textcolor[rgb]{ .31,  .506,  .741}{1}} & \multicolumn{1}{r|}{\cellcolor[rgb]{ .31,  .506,  .741}\textcolor[rgb]{ .31,  .506,  .741}{1}} & \multicolumn{1}{r|}{\cellcolor[rgb]{ .31,  .506,  .741}\textcolor[rgb]{ .31,  .506,  .741}{1}} \\
				\cmidrule{2-10}    \multicolumn{1}{l|}{Paraguay} & \cellcolor[rgb]{ .31,  .506,  .741}\textcolor[rgb]{ .31,  .506,  .741}{1} & \multicolumn{1}{r|}{\cellcolor[rgb]{ .31,  .506,  .741}\textcolor[rgb]{ .31,  .506,  .741}{1}} & \multicolumn{1}{r|}{\cellcolor[rgb]{ .31,  .506,  .741}\textcolor[rgb]{ .31,  .506,  .741}{1}} & \multicolumn{1}{r|}{\cellcolor[rgb]{ .31,  .506,  .741}\textcolor[rgb]{ .31,  .506,  .741}{1}} & \multicolumn{1}{r|}{\cellcolor[rgb]{ .31,  .506,  .741}\textcolor[rgb]{ .31,  .506,  .741}{1}} & \multicolumn{1}{r|}{\cellcolor[rgb]{ .31,  .506,  .741}\textcolor[rgb]{ .31,  .506,  .741}{1}} & \multicolumn{1}{r|}{\cellcolor[rgb]{ .31,  .506,  .741}\textcolor[rgb]{ .31,  .506,  .741}{1}} & \multicolumn{1}{r|}{\cellcolor[rgb]{ .31,  .506,  .741}\textcolor[rgb]{ .31,  .506,  .741}{1}} & \multicolumn{1}{r|}{\cellcolor[rgb]{ .31,  .506,  .741}\textcolor[rgb]{ .31,  .506,  .741}{1}} \\
				\cmidrule{2-10}    \multicolumn{1}{l|}{Peru} & \cellcolor[rgb]{ .969,  .588,  .275}\textcolor[rgb]{ .969,  .588,  .275}{2} & \multicolumn{1}{r|}{\cellcolor[rgb]{ .31,  .506,  .741}\textcolor[rgb]{ .31,  .506,  .741}{1}} & \multicolumn{1}{r|}{\cellcolor[rgb]{ .31,  .506,  .741}\textcolor[rgb]{ .31,  .506,  .741}{1}} & \multicolumn{1}{r|}{\cellcolor[rgb]{ .31,  .506,  .741}\textcolor[rgb]{ .31,  .506,  .741}{1}} & \multicolumn{1}{r|}{\cellcolor[rgb]{ .31,  .506,  .741}\textcolor[rgb]{ .31,  .506,  .741}{1}} & \multicolumn{1}{r|}{\cellcolor[rgb]{ .31,  .506,  .741}\textcolor[rgb]{ .31,  .506,  .741}{1}} & \multicolumn{1}{r|}{\cellcolor[rgb]{ .31,  .506,  .741}\textcolor[rgb]{ .31,  .506,  .741}{1}} & \multicolumn{1}{r|}{\cellcolor[rgb]{ .31,  .506,  .741}\textcolor[rgb]{ .31,  .506,  .741}{1}} & \multicolumn{1}{r|}{\cellcolor[rgb]{ .31,  .506,  .741}\textcolor[rgb]{ .31,  .506,  .741}{1}} \\
				\cmidrule{2-10}    \multicolumn{1}{l|}{Uruguay} & \cellcolor[rgb]{ .31,  .506,  .741}\textcolor[rgb]{ .31,  .506,  .741}{1} & \multicolumn{1}{r|}{\cellcolor[rgb]{ .31,  .506,  .741}\textcolor[rgb]{ .31,  .506,  .741}{1}} & \multicolumn{1}{r|}{\cellcolor[rgb]{ .31,  .506,  .741}\textcolor[rgb]{ .31,  .506,  .741}{1}} & \multicolumn{1}{r|}{\cellcolor[rgb]{ .31,  .506,  .741}\textcolor[rgb]{ .31,  .506,  .741}{1}} & \multicolumn{1}{r|}{\cellcolor[rgb]{ .31,  .506,  .741}\textcolor[rgb]{ .31,  .506,  .741}{1}} & \multicolumn{1}{r|}{\cellcolor[rgb]{ .31,  .506,  .741}\textcolor[rgb]{ .31,  .506,  .741}{1}} & \multicolumn{1}{r|}{\cellcolor[rgb]{ .31,  .506,  .741}\textcolor[rgb]{ .31,  .506,  .741}{1}} & \multicolumn{1}{r|}{\cellcolor[rgb]{ .31,  .506,  .741}\textcolor[rgb]{ .31,  .506,  .741}{1}} & \multicolumn{1}{r|}{\cellcolor[rgb]{ .31,  .506,  .741}\textcolor[rgb]{ .31,  .506,  .741}{1}} \\
				\cmidrule{2-10}    \multicolumn{1}{l|}{Venezuela} & \cellcolor[rgb]{ .31,  .506,  .741}\textcolor[rgb]{ .31,  .506,  .741}{1} & \multicolumn{1}{r|}{\cellcolor[rgb]{ .31,  .506,  .741}\textcolor[rgb]{ .31,  .506,  .741}{1}} & \multicolumn{1}{r|}{\cellcolor[rgb]{ .31,  .506,  .741}\textcolor[rgb]{ .31,  .506,  .741}{1}} & \multicolumn{1}{r|}{\cellcolor[rgb]{ .31,  .506,  .741}\textcolor[rgb]{ .31,  .506,  .741}{1}} & \multicolumn{1}{r|}{\cellcolor[rgb]{ .31,  .506,  .741}\textcolor[rgb]{ .31,  .506,  .741}{1}} & \multicolumn{1}{r|}{\cellcolor[rgb]{ .31,  .506,  .741}\textcolor[rgb]{ .31,  .506,  .741}{1}} & \multicolumn{1}{r|}{\cellcolor[rgb]{ .31,  .506,  .741}\textcolor[rgb]{ .31,  .506,  .741}{1}} & \multicolumn{1}{r|}{\cellcolor[rgb]{ .969,  .588,  .275}\textcolor[rgb]{ .969,  .588,  .275}{2}} & \multicolumn{1}{r|}{\cellcolor[rgb]{ .608,  .733,  .349}\textcolor[rgb]{ .608,  .733,  .349}{3}} \\
				\cmidrule{2-10}    \multicolumn{1}{r}{} & \multicolumn{1}{r}{} &   &   &   &   &   &   &   &  \\
				\cmidrule{2-2}      & \cellcolor[rgb]{ .31,  .506,  .741} & \multicolumn{8}{l}{Both perceptions and inequality data available} \\
				\cmidrule{2-2}      & \cellcolor[rgb]{ .969,  .588,  .275} & \multicolumn{8}{l}{Inequality was calculated with a close survey} \\
				\cmidrule{2-2}      & \cellcolor[rgb]{ .608,  .733,  .349} & \multicolumn{8}{l}{Inequality was calculated with a linear interpolation} \\
				\cmidrule{2-2}      & \cellcolor[rgb]{ .749,  .749,  .749} & \multicolumn{8}{l}{Latinobar\'ometro did not conduct survey in this year} \\
				\cmidrule{2-2}    \end{tabular}%
		\end{center}
		\begin{singlespace}  \vspace{-.5cm}
			\noindent \justify 
		\end{singlespace} 	
	}
\end{table}

\subsection{Imputation of Missing Values for the Regression Analysis}

Two of our individual-level variables (political ideology and religion) have many missing values in some country-years. To deal with this in our regressions, we imputed the average value of each variable to individuals with a missing value. In those cases, we included in the regression a dummy that takes the value one if the value of the variable was imputed and zero otherwise. The results are similar if we do not impute the values, but the sample size of the regressions is smaller.

\subsection{Comparison between Latinobar\'ometro's and SEDLAC's samples} \label{app_lb_sedlac}

To assess whether there are systematic differences between Latinobar\'ometro's sample and the household surveys' sample, in Appendix Table \ref{tab-lat-sedlac} we compare a set of variables available in both datasets during 2013. To ensure comparability across databases, we restrict the calculations to individuals over age 18 and countries with data available in both harmonization projects. 

In general, the samples are similar in observable characteristics. For instance, the average age in Latinobar\'ometro's 2013 sample is 40.6 years, while in SEDLAC it is 42.7 years. Similarly, the percentage of males is 48.9\% in Latinobar\'ometro and 47.6\% in SEDLAC. The main difference arises from educational attainment. On average, the SEDLAC sample is more educated: 46.1\% of the population has secondary education or more, while this figure is 38.8\% in Latinobar\'ometro.

\begin{table}[H]{\scriptsize
		\begin{center}
			\caption{Descriptive statistics in Latinobar\'ometro and SEDLAC, 2013 (selected countries)} \label{tab-lat-sedlac}
			\begin{tabular}{lcccccccc}
				\midrule
				& \multicolumn{2}{c}{Mean} &   & \multicolumn{2}{c}{Standard Dev.} &   & \multicolumn{2}{c}{Observations} \\
				\cmidrule{2-3}\cmidrule{5-6}\cmidrule{8-9}      &  Latinob.  &  SEDLAC  &   &  Latinob.  &  SEDLAC  &   &  Latinob.  &  SEDLAC  \\
				& (1) & (2) &   & (3) & (4) &   & (5) & (6) \\
				\midrule
				\textbf{\hspace{-1em} Panel A. Sociodemographic} &   &   &   &   &   &   &   &  \\
				Age & 40.59 & 42.68 &   & 16.43 & 17.25 &   &      14,855  &   1,004,894  \\
				Male (\%) & 48.97 & 47.63 &   & 0.50 & 0.50 &   &      14,855  &   1,004,894  \\
				Married or civil union (\%) & 56.77 & 36.41 &   & 0.50 & 0.48 &   &      14,804  &      915,117  \\
				\multicolumn{3}{l}{\textbf{\hspace{-1em} Panel B. Education and Labor market}} &   &   &   &   &   &  \\
				Literate (\%) & 91.18 & 92.17 &   & 0.28 & 0.27 &   &      14,855  &   1,004,744  \\
				Secondary education or more (\%) & 38.83 & 46.11 &   & 0.49 & 0.50 &   &      14,855  &   1,001,672  \\
				Economically active (\%) & 65.14 & 68.66 &   & 0.48 & 0.46 &   &      14,855  &   1,004,718  \\
				Unemployed (\%) & 5.78 & 4.08 &   & 0.23 & 0.20 &   &      14,855  &   1,004,718  \\
				\textbf{\hspace{-1em} Panel C. Assets and Services} &   &   &   &   &   &   &   &  \\
				Access to a sewerage (\%) & 68.76 & 63.41 &   & 0.46 & 0.48 &   &      13,799  &      975,726  \\
				Car (\%) & 26.37 & 21.09 &   & 0.44 & 0.41 &   &      11,612  &      643,350  \\
				Computer (\%) & 46.55 & 47.82 &   & 0.50 & 0.50 &   &      12,747  &      894,003  \\
				Fridge (\%) & 82.76 & 88.89 &   & 0.38 & 0.31 &   &      12,763  &      894,003  \\
				Homeowner (\%) & 74.09 & 69.64 &   & 0.44 & 0.46 &   &      14,761  &   1,003,306  \\
				Mobile (\%) & 86.91 & 91.78 &   & 0.34 & 0.27 &   &      12,754  &      896,079  \\
				Washing machine (\%) & 60.49 & 56.88 &   & 0.49 & 0.50 &   &      11,816  &      848,350  \\
				Landline (\%) & 40.22 & 39.47 &   & 0.49 & 0.49 &   &      12,736  &      896,425  \\
				\midrule
			\end{tabular}
		\end{center}
		\begin{singlespace}
			\noindent \justify \textbf{Note:} This table compares the observable characteristics of individuals in Latinobar\'ometro and SEDLAC. Summary statistics were calculated on a restricted sample (individuals aged over 18) to ensure comparability between both datasets, pooling data from 14 countries in 2013: Argentina, Bolivia, Brazil, Chile, Colombia, Costa Rica, Dominican Republic, Ecuador, El Salvador, Honduras, Panama, Peru, Paraguay, and Uruguay.
		\end{singlespace}
		
	}
\end{table}

	\clearpage	
\section{The Oaxaca-Blinder Decomposition} \label{sec_oaxaca}

\setcounter{table}{0}
\setcounter{figure}{0}
\setcounter{equation}{0}	
\renewcommand{\thetable}{C\arabic{table}}
\renewcommand{\thefigure}{C\arabic{figure}}
\renewcommand{\theequation}{C\arabic{equation}}

The starting point to decompose changes in unfairness perceptions between 2002 and 2013 is the following equation:
\begin{align}
	\text{Unfair}_{ict} = \beta_t X_{ict} + \gamma_t \text{Gini}_{ct} + \varepsilon_{ict} \quad \text{for} \quad t \in \{2002, 2013\}, 
\end{align}
where $t$ indicates the year in which perceptions are elicited and $X_{ict}$ is a vector that contains individual-level controls. The fraction of individuals who perceive the income distribution as unfair in year $t$ can be calculated as
\begin{align}
	\overline{\text{Unfair}}_{t} = \hat{\beta}_{t} \bar{X}_{t} + \hat{\gamma}_{t} \overline{\text{Gini}}_{t}  \quad \text{for} \quad t \in \{2002, 2013\}, 
\end{align}
where $\bar{X}_t$ is a vector of the average values of the explanatory variables in year $t$, and $\hat{\beta}$ the vector of OLS-estimated coefficients. The change in unfairness beliefs between 2013 and 2002 is given by
\begin{align} \label{eq-diff-fair}
	\underbrace{\overline{\text{Unfair}}_{2013} - \overline{\text{Unfair}}_{2013}}_{\equiv \Delta \text{Unfair}} = (\hat{\beta}_{2013} \bar{X}_{2013} + \hat{\gamma}_{2013} \overline{\text{Gini}}_{2013}) - (\hat{\beta}_{2002} \bar{X}_{2002} + \hat{\gamma}_{2002} \overline{\text{Gini}}_{2002})
\end{align}

Adding and subtracting $\hat{\beta}_{2002} \bar{X}_{2013} + \hat{\gamma}_{2002} \overline{\text{Gini}}_{2013}$ to equation \eqref{eq-diff-fair} yields
\begin{align} \label{eq_oaxaca}
	\Delta \text{Unfair} &=  
	\underbrace{\hat{\beta}_{2002} (\bar{X}_{2013} - \bar{X}_{2002})}_{\equiv \Delta \text{Demog.}} + 
	\underbrace{\hat{\gamma}_{2002} (\overline{\text{Gini}}_{2013} - \overline{\text{Gini}}_{2002})}_{\equiv \Delta \text{Gini}} \notag \\ &+ \underbrace{\bar{X}_{2013} (\hat{\beta}_{2013} - \hat{\beta}_{2002})+ \overline{\text{Gini}}_{2013} (\hat{\gamma}_{2013} - \hat{\gamma}_{2002})}_{\text{Residual}}
\end{align}

The first two terms of equation \eqref{eq_oaxaca} are usually known as the ``composition effect.'' These effects capture the difference between the average perceptions in 2002 and the counterfactual perceptions 2013 had the $\hat{\beta}$'s and $\hat{\gamma}$---i.e., the elasticity of perceptions to the different covariables---remained constant during the 2002--13 period. The first term captures differences in individual-level demographic variables that determine unfairness perceptions in the model (such as educational attainment, age, and employment status). The second term captures changes in aggregate trends in income inequality.

The third term of \eqref{eq_oaxaca} reflects the difference between the average fairness views in 2013 and the counterfactual fairness views in 2002 with the observable attributes of 2013. Thus, this component reflects changes in fairness views due to changes in the elasticity of the different covariables between both years. Since we cannot explain why the coefficients attached to each variable changed, this term is usually viewed as the ``unexplained'' part of the decomposition and treated as the residual of the decomposition.

\end{document}